\documentclass[useAMS,usenatbib]{mn2e}
\pdfoutput=1 
\usepackage{hyperref}
\usepackage{amsmath}
\usepackage{graphicx}
\usepackage{txfonts}
\usepackage{natbib}

\newcommand{\teff}{T_{\rm{eff}}}
\newcommand{\logg}{\log g}
\newcommand{\feh}{\rm{[Fe/H]}}

\newcommand{\ebv}{E(B-V)}

\title[$\teff, \logg$ and $\feh$ for the SkyMapper
  system]{SkyMapper stellar parameters for Galactic Archaeology on a
  grand-scale}
\author[Casagrande et al.]{\parbox{19cm}{
  L.~Casagrande$^{1,2}$\thanks{Email:luca.casagrande@anu.edu.au}, 
  C.~Wolf$^{1}$, A.~D.~Mackey$^{1,2}$, T.~Nordlander$^{1,2}$, D.~Yong$^{1,2}$ and M.~Bessell$^1$\vspace{0.2cm}}\\
  $^1$ Research School of Astronomy and Astrophysics, Mount Stromlo 
  Observatory, The Australian National University, ACT 2611, Australia\\
  $^2$ ARC Centre of Excellence for All Sky Astrophysics in 3 Dimensions
  (ASTRO 3D)}

\begin{document}

\date{Received; accepted}

\maketitle

\begin{abstract}
  The SkyMapper photometric surveys provides $uvgriz$ photometry for several
  millions sources in the Southern sky. We use DR1.1 to explore the quality of
  its photometry, and develop a formalism to homogenise zero-points across the
  sky using stellar effective temperatures. Physical flux transformations, and
  zero-points appropriate for this release are derived, along with relations
  linking colour indices to stellar parameters. Reddening-free pseudo-colours
  and pseudo-magnitudes are also introduced. For late-type stars
  which are best suited for Galactic Archaeology, we show that SkyMapper+2MASS
  are able to deliver a precision better than 100~K in effective temperatures
  (depending on the filters), $\sim 0.2$~dex for metallicities above $-2$, and
  a reliable distinction between M-dwarfs and -giants.
  Together with astrometric and asteroseismic space mission, SkyMapper promises
  to be a treasure trove for stellar and Galactic studies.   
\end{abstract}


\section{Introduction}

Photometric systems and filters are designed to be sensitive to certain spectral
features. In the case of stars, these features are driven by physical parameters
such as effective temperature,
gravity and metallicity. To accomplish this goal, filter systems are tailored to
select regions in stellar spectra where the variations of the atmospheric
parameters leave their characteristic traces with enough prominence to be
detected. A large number of photometric systems exists nowadays for different
scientific purposes \citep[e.g.,][for a review]{bessell05}, and indeed 
the advent of large scale photometric surveys is impacting every
area of astrophysics \citep[e.g.,][for a review]{ibj}.

Among the many photometric surveys is
SkyMapper\footnote{http://skymapper.anu.edu.au/}, a
$1.35$m, 32 CCDs, automated wide-field survey telescope located at Siding
Spring Observatory (Australia), undertaking a multi-epoch photometric survey of
the entire Southern sky \citep{keller07,wolf18}. The SkyMapper photometric
system builds
on the success of the $griz$ filters used by the Sloan Digital Sky Survey
\citep{fuku96,doi10}, with the added value of the $uv$ bands, designed to be
strongly sensitive to stellar parameters. The SkyMapper $u$ band mimics the
Str\"omgren $u$ filter, which covers the Balmer discontinuity and provides
good temperature sensitivity in hot stars, and gravity sensitivity across A, F
and G spectral types \citep[e.g.,][]{stro51,arnadottir10}. The SkyMapper $v$
filter is instead different from the Str\"omgren $v$, and
shifted $\sim 200\,\AA$\, towards the blue to be even more sensitive at low
metallicities, similarly to the DDO38 filter \citep{mcclure76}. The only other
existing all-sky survey measuring
intermediate $uv$ photometry is the Geneva-Copenhagen Survey \citep[GCS,][]
{nordstrom04} but at significantly brighter magnitudes than those probed by
SkyMapper, and only for FG spectral types. Nevertheless, the GCS has clearly
shown the power of intermediate Str\"omgren $uv$ photometry for Galactic
studies. Indeed, early SkyMapper data has already been very successful
at finding some of the most iron-poor stars in the Galaxy
\citep[e.g.,][]{keller14,howes16}. A full description of
the SkyMapper photometric system can be found in
\cite{bbs11}. 

In 2017, SkyMapper made available\footnote{As explained in \cite{wolf18}, the
  major improvement with respect to DR1 is a significant enhancement of the
  homogeneity of the photometric calibration. By default all queries in
  SkyMapper now return DR1.1 photometry, which is the one used in this paper.}
its Data Release 1.1 (DR1.1)
\citep{wolf18}, which provides $uvgriz$ magnitudes for over 285  million
objects across most of the southern sky ($17,200\,\rm{deg}^2$).
Although the
goal of SkyMapper is to deliver magnitudes in the AB system, when implementing
a photometric system at the telescope it is not necessarily straightforward to
adhere to the definition, and small zero-points offsets might be present.
Knowledge of these offsets is important to assess the quality of the
observations, to convert magnitudes into fluxes, as well as e.g., to compute
theoretical synthetic colours to compare with observations
\citep[e.g.,][]{cv18b,cv18}. Indeed, the first
goal of this paper is to assess the DR1.1 photometric standardization. In this
process, we develop a new method to infer photometric zero-points across the
sky, and we provide corrections to place $uvgriz$ photometry onto the AB system.

Over the next few years SkyMapper will
deliver a uniquely powerful dataset to investigate stellar populations across
the Galaxy, enabling studies in most areas of Galactic Archaeology. 
Thus, the second goal of this paper is to derive empirical calibrations relating
basic stellar parameters ($\teff, \logg$ and $\feh$) to SkyMapper photometry.
Stellar effective temperatures are derived implementing SkyMapper photometry
into the InfraRed Flux Method \citep[IRFM,][]{c10}. The sensitivity of
SkyMapper photometry to $\feh$ and $\logg$ is explored using over a quarter of
a million stars in common between SkyMapper and the spectroscopic GALactic
Archaeology with Hermes survey \citep[GALAH,][]{buder18}. We are able to
compare photometric $\teff$ from SkyMapper to spectroscopic ones from GALAH,
as well as to explore the sensitivity of SkyMapper filters to stellar
parameters. This exercise goes beyond the importance of cross-validating the
two surveys. In fact, SkyMapper is ultimately expected to be magnitude-complete 
down to $g \simeq 22$, thus reaching several magnitudes fainter than
GALAH, and approximately the same magnitude limit as {\it Gaia}, greatly
enlarging the volume within which we can do Galactic Archaeology. The
complementarity of
SkyMapper to {\it Gaia} is enormous, especially at this stage when $BP$ and
$RP$ spectra have not been released yet, meaning that {\it Gaia} stellar
parameters are based only on $G_{\rm{BP}}, G$ and $G_{\rm{RP}}$ photometry, and
thus subject to strong assumptions and degeneracies.

\begin{figure*}
\begin{center}
\includegraphics[scale=0.47]{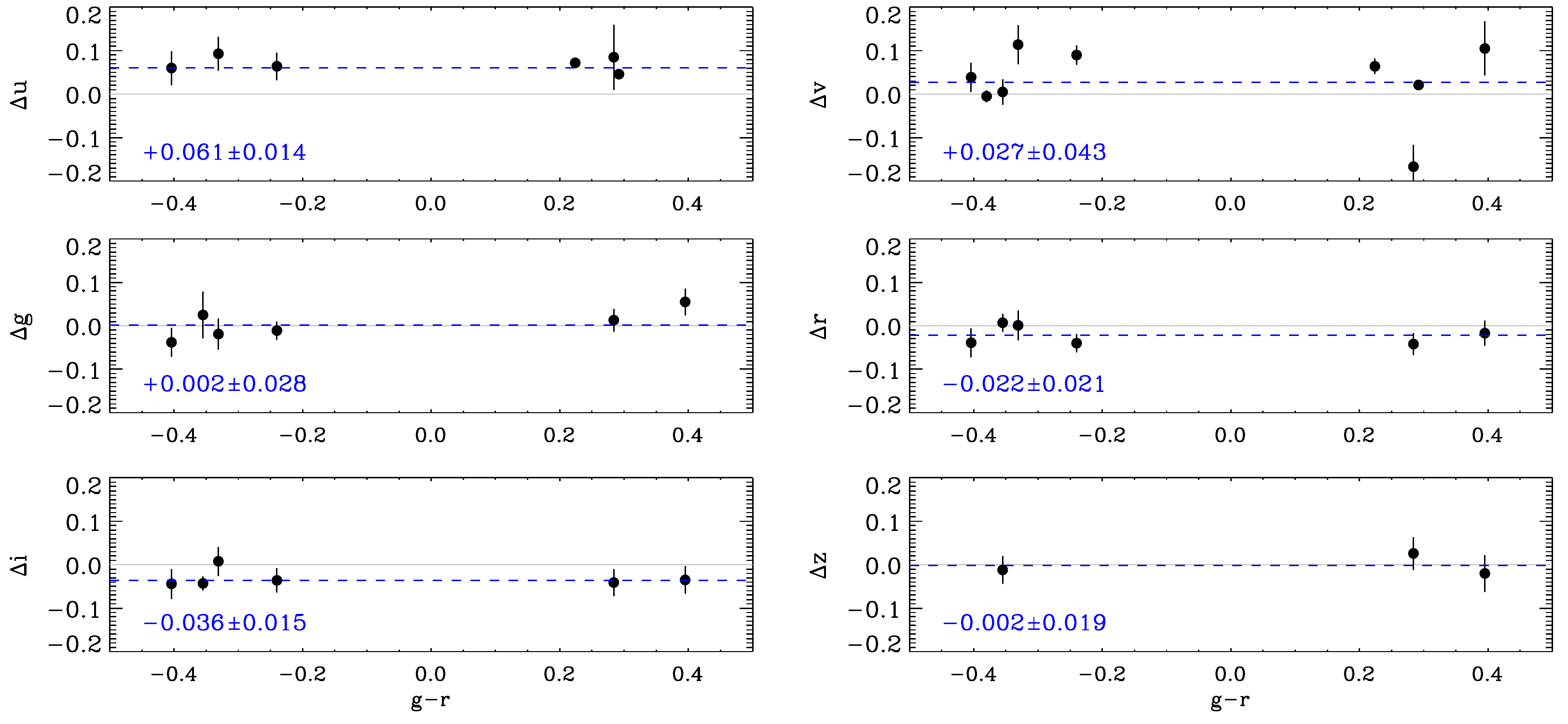}
\caption{Observed SkyMapper minus AB magnitudes computed for stars in the
  CALSPEC library as function of their $g-r$ colour. For each band, only stars
  with no SkyMapper flags and no source within 15 arcsec have been retained.
  $\epsilon_{\zeta}$ are shown by dashed lines, with the weighted difference
  $\pm$ the square root of the weighted sample variance indicated at the
  bottom of each panel.}\label{fig:zp1}
\end{center}
\end{figure*}

\section{The SkyMapper system}\label{sec:sms}

A source having flux $f_{\lambda}$ and observed through a system response
function $T_\zeta$ (which includes the total throughput reaching the observer
over the bandpass $\zeta$) will have an AB magnitude \cite[see e.g.,][for the
  photon-counting formalism adopted here]{bm12,cv14}:
\begin{equation}\label{eq:AB}
  m_{\zeta, \tt AB}= -2.5 \log \frac{\int_{\nu_i}^{\nu_f} f_{\nu} T_\zeta \rm{d}\ln\nu}{f_{\nu}^{0} \int_{\nu_i}^{\nu_f} T_\zeta \rm{d}\ln\nu} = 
  -2.5 \log \frac{\int_{\lambda_i}^{\lambda_f} \lambda f_{\lambda}
 T_\zeta \rm{d}\lambda}
{f_{\nu}^{0} c \int_{\lambda_i}^{\lambda_f}
 \frac{T_\zeta}{\lambda} \rm{d}\lambda},
\end{equation}
where $f_{\nu}^{0}=3.631 \times 10^{-20} \rm{erg\,s^{-1}\,cm^{-2}\,Hz^{-1}}$ and
$c$ is the speed of light. 
The actual realization of a photometric system at the telescope is far from
the trivial definition given above. More often than not, zero-point
corrections $\epsilon_\zeta$ are needed in each band to adhere to the
definition \citep[e.g.,][for the SDSS
  system]{eisen06,hb06}. Thus, it is worth checking whether this is also the
case for SkyMapper.
Currently, each SkyMapper exposure is standardized as closely as possible
to the AB system through comparison with APASS and 2MASS photometry
\citep{apass2016,skr06}.
SkyMapper standardized magnitudes ({\tt SM}) can thus be written:
\begin{equation}\label{eq:SK}
m_{\zeta, \tt SM} = m_{\zeta, \tt AB} + \epsilon_{\zeta},
\end{equation}
where $\epsilon_\zeta$ allows for possible departure from the AB definition. In
the most general form, these departures could depend on various factors such as
position across the sky, magnitudes or colours. These effects are explored
and discussed later in the paper.

Here, for each SkyMapper filter ($\zeta = u,v,g,r,i,z$) we adopt the system
response functions reported in \cite{bbs11}. Since the SkyMapper system
response functions are well characterized, one way of determining
$\epsilon_\zeta$ is to use measured
absolute spectrophotometry (i.e. $f_{\lambda}$) to compute $m_{\zeta, \tt AB}$ via
Eq.~(\ref{eq:AB}). Comparison with observed SkyMapper magnitudes allows then to
determine $\epsilon_{\zeta}$ via Eq.~(\ref{eq:SK}). The HST
CALSPEC\footnote{http://www.stsci.edu/hst/observatory/crds/calspec.html}
library offers the most accurate absolute spectrophotometry available to date,
which is of order of a few percent, or better for stars with STIS/NICMOS
observations \citep{bdc01,bo07,bo14}. We remark that a systematic uncertainty
of order 1 percent in absolute flux translates into
$2.5\log(1.01) \simeq 0.01$~mag zero-point uncertainty. We compute photometric
errors by taking into account systematic and statistical errors as reported
for each absolute flux in CALSPEC. For STIS/NICMOS observations the impact of
statistical errors is usually smaller, as they mostly compensate over a
bandwidth. We find 11 stars in SkyMapper that also have CALSPEC absolute
spectrophotometry, and Figure \ref{fig:zp1} shows the difference between the
magnitudes observed and those computed via Eq. (\ref{eq:AB}). 

From Figure \ref{fig:zp1}, $u$ is the only band displaying a $>3\sigma$ offset
from the AB system. $v$ also seems to be offset, but with a large scatter, the
weighted difference and weighted sample variance not changing significantly if 
we were exclude the biggest outlier ($0.030 \pm 0.034$ instead of
$0.027 \pm 0.043$). $g$ and $z$ are
consistent with being on the AB system, whereas small offsets are present for
the $r$ and $i$ band, but those are only marginally significant (around 1 and
2$\sigma$, respectively). The minimal offset and typical $0.02$ mag
scatters for the
$griz$ filters support the conclusion of \cite{wolf18}, who found a scatter of
2 percent with respect to the AB photometry from Pan-STARRS1. In Figure
\ref{fig:zp1}, the error bars of most points reach the zero-point corrected
dashed-lines, except for $v$ band. This band is also characterized by a rather
large scatter, which warrants further investigation. As we discuss in
the next Section, a larger number of spectrophotometric standards across
the sky would be necessary to draw a firmer conclusion. With this goal in mind,
in the next Section we develop a new method to derive robust photometric
zero-points using a cohort of stars across the sky.

\section{Photometric zero-points from the absolute $\teff$ scale}\label{sec:irfm}

In this section we explore an alternative approach to
derive photometric zero-points for the SkyMapper system. To do so, we use the
IRFM, which
provides a nearly model independent and elegant technique for determining
stellar effective temperatures \citep[e.g.,][]{blackwell79,blackwell80}. The
IRFM relies  on the ratio between the bolometric ($\mathcal{F}_{\rm{bol}}$) and
the infrared monochromatic flux ($\mathcal{F}_{\rm{IR}}$) of a star measured on
the Earth. Both quantities are determined observationally. This ratio is
compared to the one defined on a stellar surface element, i.e. the bolometric
flux $\sigma \teff^4$ and the theoretical surface infrared monochromatic flux:
\begin{equation}\label{eq:IRFM}
  \frac{\mathcal{F}_{\rm{bol}}\rm{(Earth)}}{\mathcal{F}_{\rm{IR}}\rm{(Earth)}} =
  \frac{\sigma \teff^4}{\mathcal{F}_{\rm{IR}}{\rm(model)}}.
\end{equation}
When working in the Rayleigh-Jeans tail, the model infrared flux is
largely dominated by the continuum and relatively easy to compute, with a
roughly linear dependence on $\teff$ and very little affected by other stellar
parameters, such as metallicity and surface gravity \citep[as extensively
  tested in the literature, e.g,][]{blp91,alonso96:irfm,c06}. The problem is
therefore reduced to a proper derivation of stellar fluxes, and once this is
done Eq.~(\ref{eq:IRFM}) can be rearranged to return its only unknown: $\teff$.
The implementation we adopt for the IRFM uses multi-band optical and
infrared photometry to recover $\mathcal{F}_{\rm{bol}}$ and
$\mathcal{F}_{\rm{IR}}$. An iterative procedure in $\teff$ is adopted to cope
with the mild 
dependence on stellar parameters of the flux outside photometric bands (i.e.,
the bolometric correction), and of the theoretical surface infrared
monochromatic flux. More specifically, for each star, we
interpolate over a grid of synthetic model fluxes, starting with an initial
estimate of the stellar effective temperature, and working at fixed $\feh$ and
$\logg$ until convergence is reached in $\teff$. Further details can be found in
\cite{c06,c10}. In essence, the method relies on a proper derivation
of physical fluxes ($\rm{erg}\,\rm{s}^{-1}\,\rm{cm}^{-2} \rm{\AA}^{-1}$) from
magnitudes, meaning that the IRFM strongly depends on the absolute calibration
underlying the photometric systems used into it. Without exaggeration, this is
the most critical point when implementing the method \citep[e.g.,][]{bpa90}.
\cite{c10} further highlighted how differences among IRFM scales in the
literature
can be simply explained by changing the absolute calibration of the adopted
photometric systems, or equivalently using different photometric zero-points.
This means that if we have a set of stars for which we accurately know their
effective temperatures, we can implement a given photometric system (SkyMapper
in this case) into the IRFM, and modify the adopted photometric zero-points
until we are able to reproduce known effective temperatures.

As we have already discussed, the adopted implementation of the IRFM relies on
multiband optical and infrared photometry to recover the bolometric flux. The
infrared monochromatic flux is derived using only infrared magnitudes (2MASS
$JHK_s$ in this case). The infrared absolute calibration and zero-points have
already been determined in \cite{c10} via solar-twins, and are kept unchanged
here. An in-depth discussion of the flux associated to each SkyMapper magnitude
is provided in the Appendix. For the sake of applying the IRFM,
here it suffices to say that for each star we always require having 2MASS
$JHK_s$ magnitudes (with combined photometric errors $<0.15$ mag), plus at
least one SkyMapper band. A band is used only if it
has no SkyMapper flag, and no source within 15 arcsec. We also apply a
threshold on photometric errors, $<0.1$ magnitude for $u$ and $v$ band, and
$<0.04$ for $griz$, as we discuss in Section \ref{app:zp}.

To summarise, in our method for each star we input measured values of
$\logg$, and $\feh$, observed magnitudes (and reddening if present) to
derive $\teff$ via Eq.~(\ref{eq:IRFM}). Converting observed magnitudes into
fluxes introduces the dependence on photometric zero-points, and link them to
a physical quantity
such as the stellar effective temperature. The dependence on synthetic stellar
fluxes is needed to derive bolometric corrections, but besides this, at no
point we make use of theoretical predictions between magnitudes and colours.
Empirical colour-$\teff$ relations can be easily derived from the IRFM, as we
do later on in Section \ref{sec:colorte}. Once these relations are available,
one could use them to link photometric zero-points to stellar effective
temperatures bypassing the IRFM. While viable, we have not explored this
approach, as it would introduce the extra ladder of building these relations.

\begin{table}
\centering
\caption{Average photometric zero-points $\epsilon_{\zeta}$ and characteristic
  parameters of the SkyMapper system.}\label{tab:zp}
\begin{tabular}{cccccc}
\hline 
  &  $\epsilon_{\zeta}$  &  $G(\lambda)$  &  $H(\lambda)$ & $Bw(\lambda)$ & $\lambda_{\rm{eff}}$ \\
  &   &  $[\rm{cm}^{-1} \AA^{-1}]$  &  $[\rm{cm}^{-1}]$  &  $[\AA]$  & $[\AA]$ \\
\hline            
  $u$  &  $+0.032 \pm 0.020$ &    $8.086$     & $3446.6$ & $426.2$  & 3537 \\
  $v$  &  $+0.033 \pm 0.022$ &    $6.796$     & $2168.4$ & $319.1$  & 3874 \\
  $g$  &  $+0.009 \pm 0.014$ &    $3.882$     & $5631.8$ & $1450.6$ & 5016 \\
  $r$  &  $+0.006 \pm 0.010$ &    $2.654$     & $3752.8$ & $1414.1$ & 6078 \\
  $i$  &  $-0.012 \pm 0.008$ &    $1.657$     & $2065.3$ & $1246.2$ & 7734 \\
  $z$  &  $-0.001 \pm 0.006$ &    $1.195$     & $1385.0$ & $1158.6$ & 9121 \\
  \hline
\end{tabular}
\begin{minipage}{0.48\textwidth}
  $\epsilon_{\zeta}$ are those derived from Figure \ref{fig:zp2}, and must be
  subtracted from SkyMapper photometry to reproduce the AB system.
  $Bw(\lambda)$ is the bandwidth of the filters, whereas $G(\lambda)$ and
  $H(\lambda)$ are attributes necessary to derive monochromatic and
  in-band physical fluxes (see discussion in the Appendix). The
  spectrum of Vega has been adopted to compute the effective wavelength
  $\lambda_{\rm{eff}}$.
  Note that while $\epsilon_{\zeta}$ are specific for DR1.1, $G(\lambda)$,
  $H(\lambda)$, $Bw(\lambda)$ and $\lambda_{\rm{eff}}$ are valid for any future
  SkyMapper release (unless filter transmission curves are revised). 
\end{minipage}
\end{table}

\begin{figure*}
\begin{center}
\includegraphics[scale=0.47]{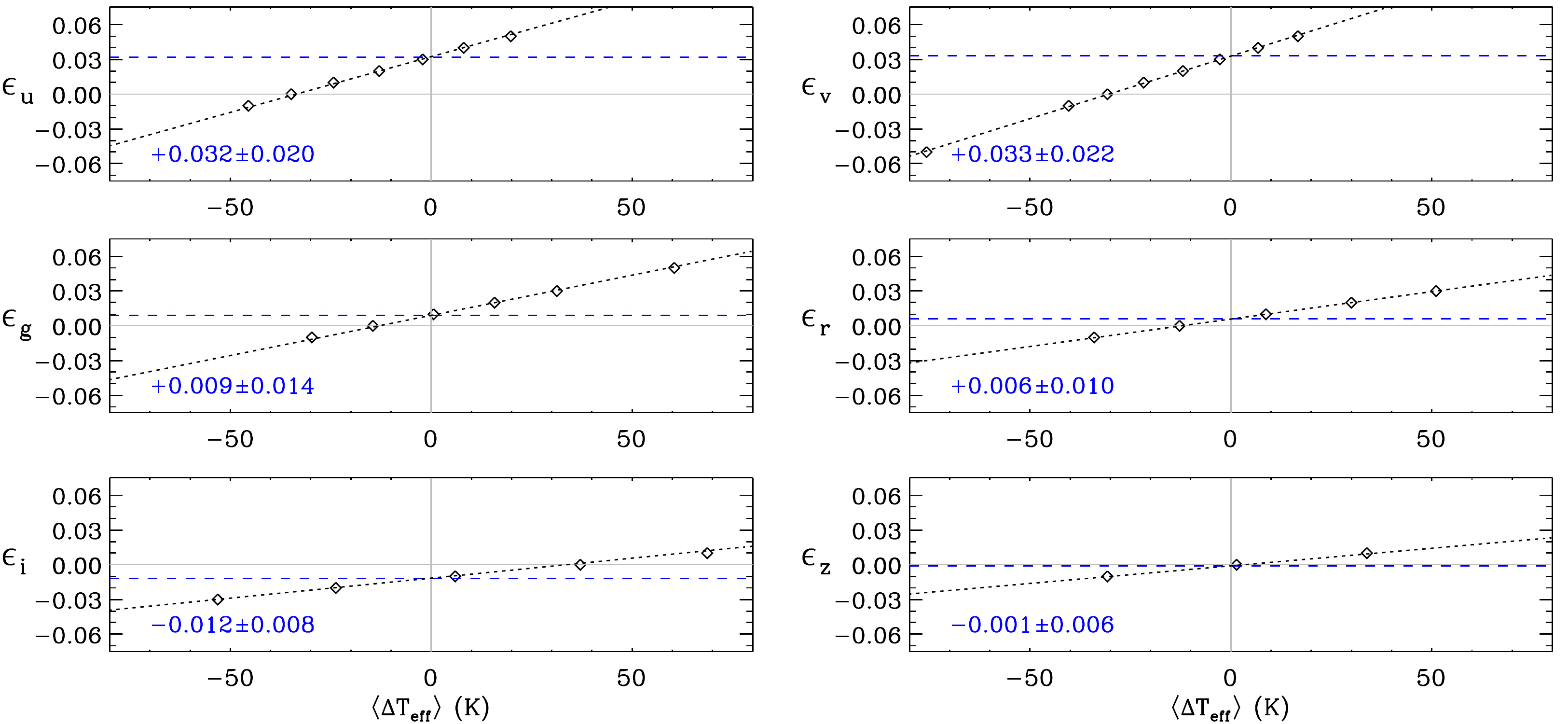}
\caption{
  Photometric zero-points determined via the effective temperature scale.
  Diamonds are the weighted average of the effective temperature difference
  (SkyMapper$-$Reference sample) when SkyMapper zero-points are varied in the
  IRFM. Dotted lines are linear fits to the
  points. The adopted $\epsilon_\zeta$ (dashed blue lines) are determined from
  the intersection of the dotted lines with
  $\left< \Delta \teff \right>=0$, and they are indicated at the bottom of each
  panel. See text for further details.}\label{fig:zp2}
\end{center}
\end{figure*}

\subsection{Reference Sample}\label{sec:refsam}

As we have explained in the previous Section, in order to infer the SkyMapper
DR1.1 photometric zero-points we need a sample of stars for which we
accurately know their $\teff$. To this purpose we use stars from
\cite{c10,c11} whose effective temperatures were homogeneously determined via
the IRFM, and for which the uncertainty on the zero-point of the $\teff$ scale
is of order $20$~K. This accuracy implies that we are able to pin down
photometric zero-points to about $0.01$~mag. In order for this exercise
to be entirely differential in $\teff$, for each star we adopt parameters
identical to \cite{c10,c11}, i.e. the same $\feh$, $\logg$, 2MASS photometry,
and reddening (usually absent, or very small due to the nearby
nature of the sample). We also remark that for our purposes it is essential to
have stars from a well controlled sample, or systematic differences between
heterogeneous $\teff$ scales (e.g., using literature compilations) would
dominate over the zero-point effects we wish to determine. Crucially, the
zero-point of the effective temperature scale will impact the absolute flux
scale, and hence the $\epsilon_\zeta$ we derive. Stars with reliably
measured angular diameters would provide an equally good reference set
\citep[e.g.][]{karo,twhite}, but only a handful of such objects are presently
available, and because of their
brightness they are also saturated in SkyMapper. We remark that the $\teff$
scale we adopt has been tested against interferometric angular diameters
confirming its accuracy \citep{c14b,karo,twhite}.

We find a total of $544$ stars having a measurement in at least one SkyMapper
band, and effective temperatures from \citet[][which we refer to as the
  ``Reference sample'']{c10,c11}. When $\teff$ are determined
implementing SkyMapper photometry into the IRFM, we refer to the same stars as
the ``SkyMapper sample''. While nearly all $544$ stars in this sample have
$uv$ photometry, only a small percentage have $griz$ measurements — the number
of available stars in these passbands varies between 19 and 32. This is due to
the fact that stars in \cite{c10,c11} are quite bright, and the saturation
limit for $griz$ is brighter than for $uv$ magnitudes.
\begin{figure*}
\begin{center}
\includegraphics[scale=0.8]{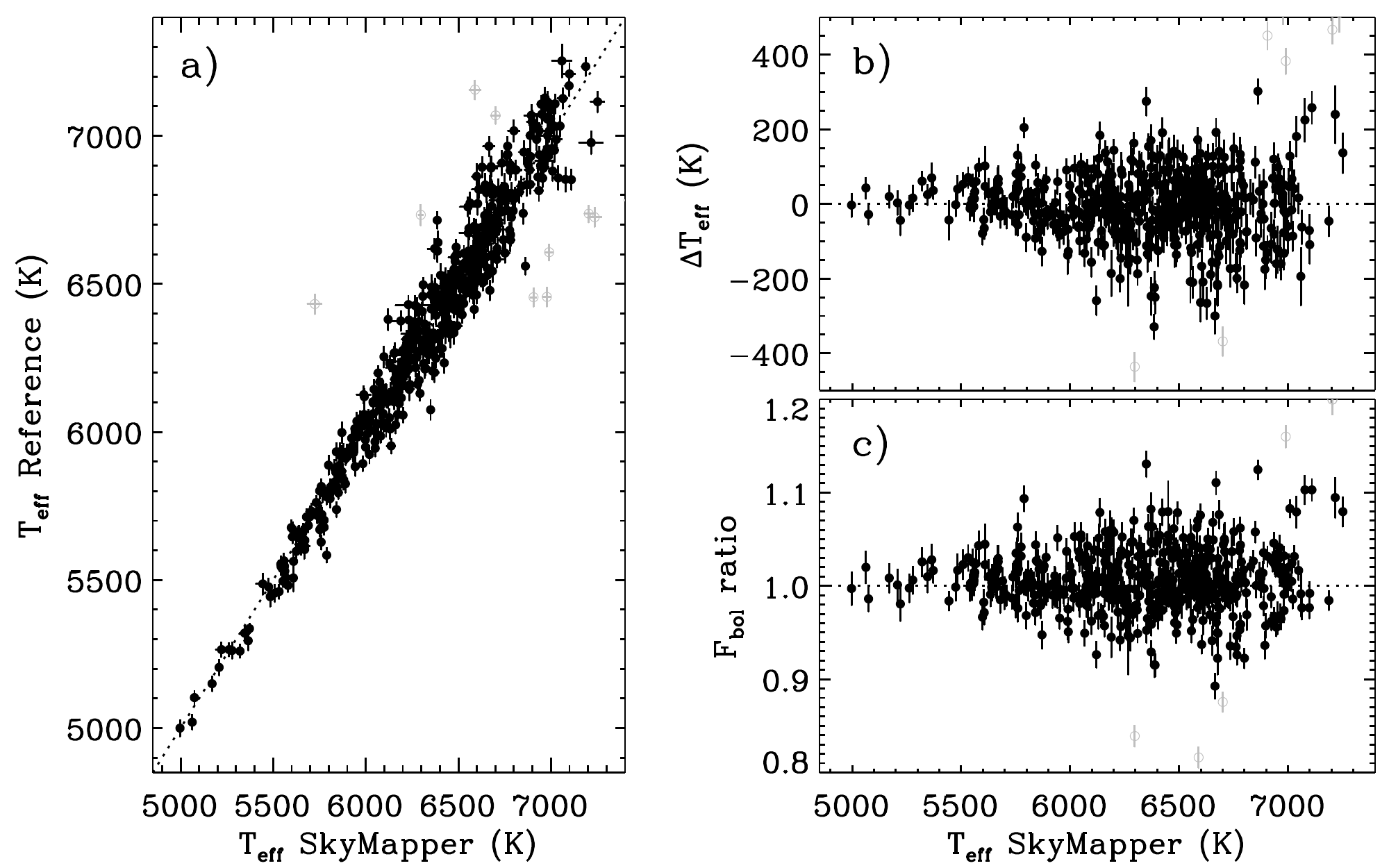}
\caption{Panel a): comparison between the effective temperatures obtained
  implementing SkyMapper photometry into the IRFM (with zero-points reported in
  Table \ref{tab:zp}), and the Reference sample of Casagrande et al. (2010,
  2011). Panel b): effective temperature difference
  (SkyMapper$-$Reference). Panel c): relative difference in bolometric
  flux (SkyMapper/Reference) for the same stars. Error bars are internal
  uncertainties obtained from a Monte Carlo simulation on photometric errors.
  For each star we use as many SkyMapper bands as possible, depending
  on quality flags and photometric errors. Stars marked in grey have been
  removed with a $3\sigma$ clipping. See text for details.}
\label{fig:cal}
\end{center}
\end{figure*}

\subsection{$uvgriz$ zero-point determination}\label{sec:zp}

We implement the IRFM using one SkyMapper band at the time (in addition to
2MASS, which is always used), and vary its $\epsilon_\zeta$ across a
suitable range, until on average stars in the SkyMapper sample have the same
$\teff$ as in the Reference sample i.e., we reproduce the zero-point of our
adopted temperature scale. This is done computing $\left< \Delta \teff \right>$,
which is the weighted average of the effective temperature difference between
the SkyMapper and the Reference sample. For stars in both samples, weights are
given by internal $\teff$ uncertainties: we run a Monte Carlo simulation into
the IRFM to assess the degree to which effective temperatures are affected by
the photometric uncertainties in the input data. For photometric errors beyond
$0.04$ mag in $griz$, we note a slight correlation with $\Delta \teff$, whereas
we do not see any for $uv$ bands (whose maximum photometric errors are around
$0.1$ mag). Hence, when computing $\left< \Delta \teff \right>$ we exclude
stars with errors larger than the values quoted above. We also apply a
3$\sigma$ clipping to remove stars with large effective temperature
differences, and we track down the reason of those in the next Section. 

The zero-point of the SkyMapper $\teff$ scale varies linearly with the value
assumed for each $\epsilon_{\zeta}$ into the IRFM. This means that the correct
value to adopt for $\epsilon_{\zeta}$ can be determined by a linear fit
intersecting an average effective temperature difference of zero. This is
shown in Figure \ref{fig:zp2}, and the zero-points so derived are reported in
Table \ref{tab:zp}. Uncertainties are obtained by adding to the
uncertainty of the intercept, the systematic if the reference $\teff$ scale
were to be shifted by $\pm 20$~K (which is the zero-point uncertainty of the
Reference sample). We remark that the zero-points we determine in this way are
usually
in good agreement with those obtained from the
CALSPEC spectrophotometry. The largest discrepancy is only $1.4\sigma$, and
the sign of the zero-points agrees for all, but $r$ band (compare Figure
\ref{fig:zp1} with Figure \ref{fig:zp2}). 

The zero-points in Table \ref{tab:zp} must be subtracted from the SkyMapper
DR1.1 magnitudes if one wishes to place them onto the AB system (or
conversely, they must be added to the AB definition to replicate SkyMapper
DR1.1 magnitudes).
Importantly, these zero-points are global. We discuss in the next Section
their dependence (or lack thereof) on sky-position and magnitudes. 

With the zero-points appropriate for each $uvgriz$ filter, we can then apply the
IRFM using as many SkyMapper bands as possible. Figure
\ref{fig:cal} confirms that when using more SkyMapper bands in the IRFM we
still reproduce the effective temperature scale of the Reference sample (as
one would expect), the weighted difference being $0 \pm 2$~K, with an rms of
$88$~K. There are some clear outliers, which stem from spatial variations of
zero-points across the sky (see Section \ref{app:zp}). Notice that although we
have discussed everything in terms of $\teff$, by changing the SkyMapper
zero-points we are also able to reproduce on average the same bolometric
fluxes (and thus angular diameters) of the Reference sample; the weighted
ratio of bolometric fluxes agrees to $0.5 \pm 0.1$, with a 3 percent rms.
(Figure \ref{fig:cal}c). The above differences would be $-31$~K and $-0.9$
percent in flux if no zero-points were applied (i.e., wrongly assuming perfect
standardization to the AB system) and $-5$~K and $0.25$ percent in flux if
using the zero-points determined from the CALSPEC spectrophotometry.

\subsection{Spatial dependence of SkyMapper zero-points}\label{app:zp}

\begin{figure*}
\begin{center}
\includegraphics[scale=0.5]{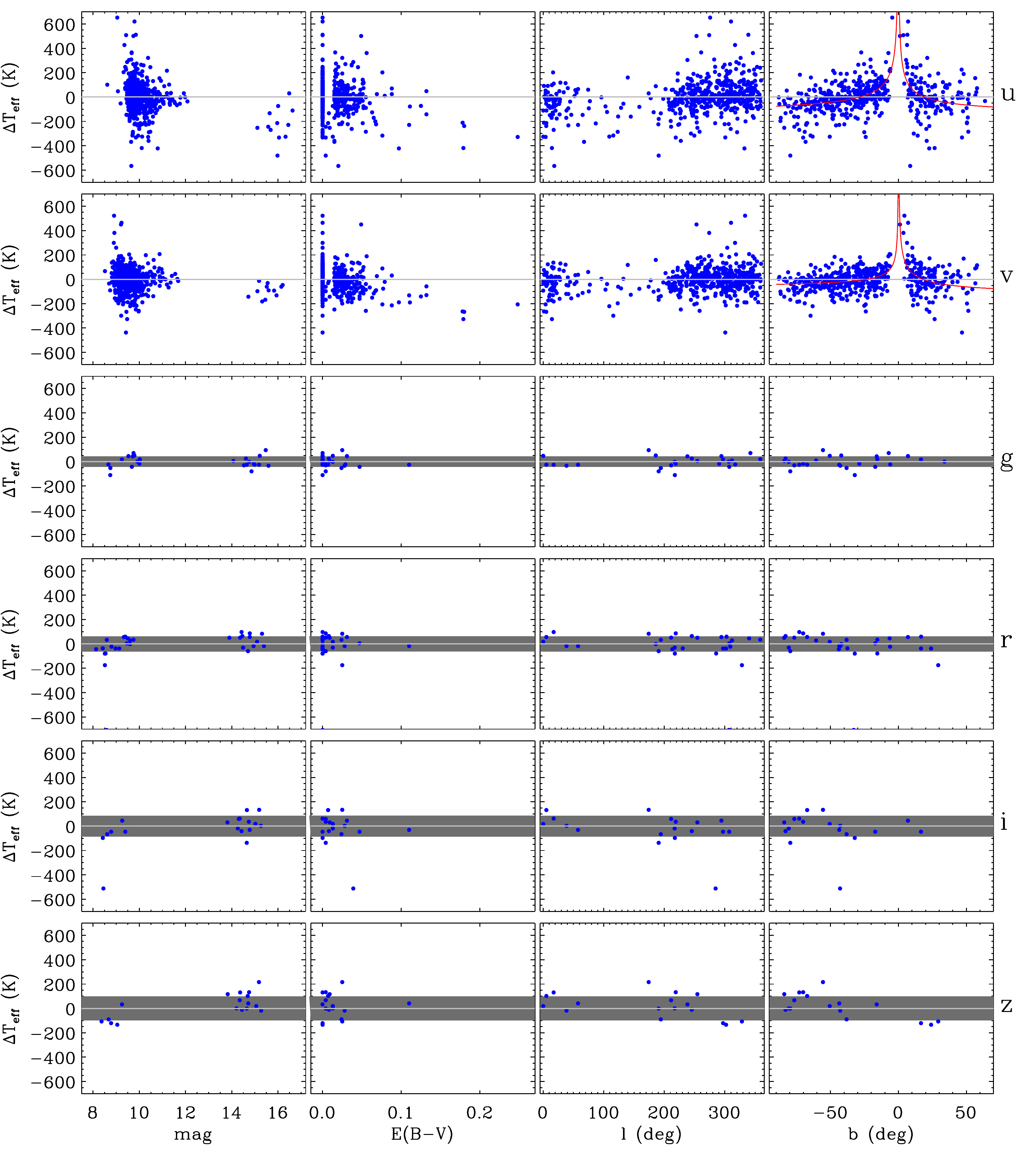}
\caption{From top to bottom: $\Delta\teff$ (SkyMapper$-$Reference) for $uvgriz$
  bands, as function of magnitudes, reddening, Galactic longitude ($l$) and
  latitude ($b$). Dark grey areas correspond to $\teff$ variations of $0.03$
  mag in $griz$, which amount to the scatter and mean offset reported for
  those bands by \citet{wolf18}. Red lines for $u$ and $v$ bands are a fit of
  $\Delta\teff$ versus $b$.}
\label{fig:spatial}
\end{center}
\end{figure*}

Ideally, photometric zero-points should be the same for all stars in the sky,
independently of anything else. However, there can be a number of reasons why
this assumption breaks down \citep[see e.g.,][for a sobering discussion on
  the difficulty of standardizing observations]{Stetson05}. The method presented
in Section \ref{sec:irfm} to determine SkyMapper zero-points has the advantage
that it can be applied to a large sample of stars (instead of the handful
having CALSPEC spectrophotometry), and thus it can be used to explore the
dependence of photometric zero-points on various parameters. This is done in
Figure \ref{fig:spatial}, which shows the effective temperature difference
(SkyMapper$-$Reference) when applying the zero-points of Table \ref{tab:zp},
and running one SkyMapper band at the time in the IRFM. 

\begin{figure*}
\begin{center}
\includegraphics[scale=0.5]{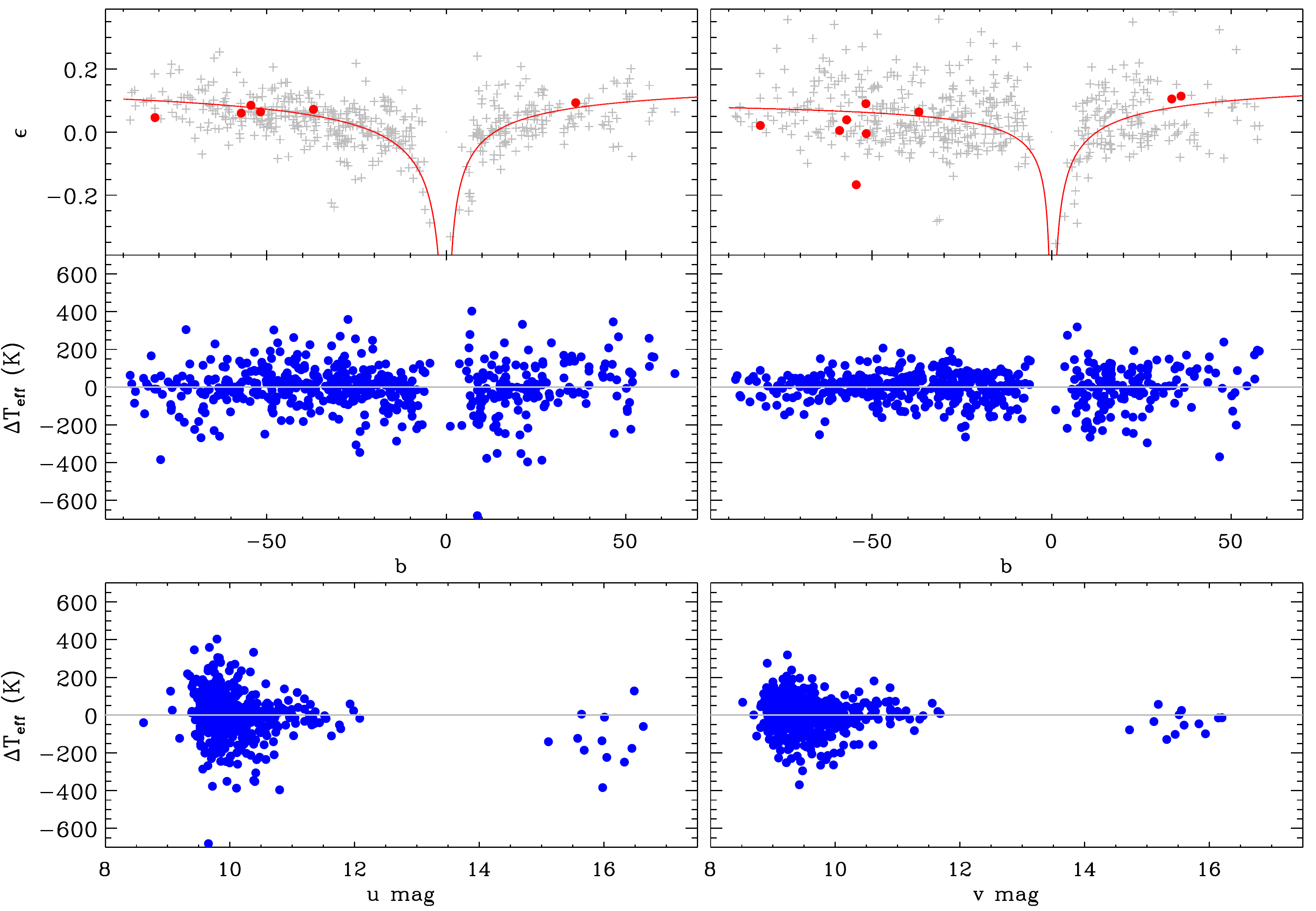}
\caption{Top panels: continuous line shows the $u$ (left) and $v$ (right)
  zero-point dependence on Galactic latitude ($b$), as per Eq.~(\ref{eq:ugb})
  and (\ref{eq:vgb}). Filled circles are the observed minus AB magnitudes for
  CALSPEC stars. Crosses are the difference between SkyMapper and Str\"omgren
  $u$ (left) and $v$ (right) magnitudes, as explained in the text.
  Middle and lower panels show $\Delta\teff$ (SkyMapper$-$Reference) as
  function of $b$ and magnitudes, after correcting SkyMapper zero-points.}
\label{fig:zpb}
\end{center}
\end{figure*}

While only a handful of points are available for $griz$ bands, no obvious trends
can be found. Using the linear mapping of Figure \ref{fig:zp2} between
zero-point shifts and $\Delta\teff$, we convert the $\sim 0.03$ magnitude scatter
reported by \cite{wolf18} for $griz$ filters into an effective temperature
scatter (grey bands in Figure \ref{fig:spatial}). Most of the points are
consistent with the location of the grey bands, thus confirming the conclusion of
\cite{wolf18}. However, large scatter and spatial trends are observed for $u$
and $v$ band, suggesting that the zero-points of those two bands are not
standardized as well as for the other SkyMapper filters. This
is not entirely unexpected, considered that SkyMapper does not observe $uv$
standards.

These trends are very clear as function of Galactic latitude $b$, although they
also appear in Galactic longitude, RA and declination because of correlation
among coordinates. To try understanding their origin, we briefly recall how
the photometric calibration is achieved in SkyMapper, and refer to
\cite{wolf18} for
further details. Instead of using standard stars, photometric zero-points to
standardize instrumental $u$ and $v$ magnitudes are estimated using
transformations which involve APASS $g$ magnitudes, a dereddened colour term,
and a reddening estimate. The dereddened colour term comes from converting
APASS magnitudes into Pan-STARRS1, and then a linear Pan-STARRS1 to SkyMapper
relation derived from unreddened stellar templates. The reddening
estimate is based on a rescaling of the \cite{sfd98} map. This procedure
defines the average zero-points for each frame. The actual zero-points applied
to each star come from fitting the differences in the predicted (from the
above procedure) and instrumental magnitudes for each star as function of
spatial position on the CCDs. This is done to take into account atmospheric
extinction
gradients across the large field of view of the SkyMapper telescope. This
approach proves to work remarkably well for $griz$ bands, as confirmed by the
$2.3$ percent scatter (and up to $1$ percent mean offset) for stars in common
between Pan-STARRS1 and calibrated SkyMapper magnitudes \citep[see]{wolf18}.
However, a higher scatter is to be
expected in $uv$ bands, because of their stronger sensitivity to stellar
parameters, and reddening. In fact, the above procedure to predict $uv$
magnitudes for a given star has a dispersion of order $0.1$~mag or more.
However, assuming only random errors, the formal uncertainty on the $uv$
zero-points is often well below $0.01$~mag, because zero-points are typically
determined using several hundred stars in each frame. Nevertheless, Figure
\ref{fig:spatial} suggests that the quality of $uv$ magnitudes is
substantially poorer than the percent level achieved for $griz$. The strong
$\Delta \teff$
trend as function of Galactic latitude likely
stems from the reddening prescriptions adopted to calibrate SkyMapper
magnitudes, as described above. In fact, $\Delta \teff$ grows positive and
larger closer to the plane, meaning that $\teff$ ($uv$ magnitudes) in
SkyMapper are overestimated (too bright) close to the plane, and vice versa at
high Galactic latitudes. Since we adopt the same reddening for the Reference
and the SkyMapper sample (and reddening for these stars is typically very low,
see discussion in Section \ref{sec:refsam}) this can only mean that the $uv$
zero-points adopted to standardize DR1.1 magnitudes are overcorrected for
reddening close to the plane, and vice versa at high latitudes.

In Figure \ref{fig:spatial} we fit $\Delta\teff$ as function of $1/\sqrt{|b|}$,
and use the mapping of Figure \ref{fig:zp2} to derive how zero-points vary
across the sky. We obtain the following functional forms:
\begin{equation}\label{eq:ugb}
\epsilon_u = \left\{\begin{array}{ll}
0.198 - 0.727/\sqrt{b}   & b>0^\circ\\
0.198 - 0.886/\sqrt{|b|} & b<0^\circ
\end{array}\right.
\end{equation}
and
\begin{equation}\label{eq:vgb}
\epsilon_v = \left\{\begin{array}{ll}
0.200 - 0.710/\sqrt{b}   & b>0^\circ\\
0.125 - 0.451/\sqrt{|b|} & b<0^\circ
\end{array}\right.
\end{equation}
where these zero-points must be subtracted from SkyMapper magnitudes to
reproduce
the AB system, and $b$ is the Galactic latitude in degrees. The lines
in the top panels of Figure \ref{fig:zpb} show the dependence of these zero-points
on Galactic latitude. We also show the
zero-points as traced by
CALSPEC standards (filled circles), as well as the difference between SkyMapper
and Str\"omgren $u$ and $v$ magnitudes for stars in the GCS (grey crosses). In
comparing with
Str\"omgren photometry, an arbitrary shift is applied to bring the grey
crosses onto the continuous line, since Str\"omgren photometry is not onto
the AB system. We remark that in no instance Str\"omgren $u$ and $v$
magnitudes were used to derive $\teff$ for our stars, yet the same trend is
found as function of Galactic latitude. This is particularly clear for $u$,
where the SkyMapper and Str\"omgren transmission curves are nearly identical,
whereas the SkyMapper $v$ band is shifted $\sim 200\,\AA$\, towards the blue
compared to the Str\"omgren one. Applying our zero-point corrections to
SkyMapper
magnitudes removes the major trend in $\Delta\teff$ versus
$b$ (middle panels). The trends with $u$ and $v$ magnitudes seen in Figure
\ref{fig:spatial} are also largely corrected for, and although not shown, the
fit as function of $b$ is sufficient to remove the wobbling trends with Galactic
longitude, as well as RA and declination.
We have previously described how the $uv$ standardization is done in DR1.1, and
pointed to reddening as the main cause for zero-point variations across
the sky. Although other systematic effects might still remain, we prefer
to have a minimum number of parameters in Eq.~(\ref{eq:ugb}) and
Eq.~(\ref{eq:vgb}). Our fits remove the main trend as function of $b$, albeit
in Figure \ref{fig:zpb} $u$ and $v$ have still a scatter of $120$~K and
$90$~K, respectively (for comparison, the scatter in the SkyMapper other bands
is between $50$ and $100$~K). These translate to photometric uncertainties of
order $0.1$~mag for $u$ and $v$. Interestingly though, the scatter when
comparing $\teff$ obtained implementing $u$ and $v$ band into the IRFM is much
smaller, $65$~K, which implies an uncertainty of order $0.06$~mag in $u-v$.
This likely indicates a degree of correlation between these two bands, which is
not surprising given the similar standardization procedure in DR1.1 for the two
filters.

Our proposed zero-point corrections amount to roughly $\pm0.1$mag across the
sky, except for regions close to the Galactic plane. We remark that we have a
handful of stars with $|b|<10^\circ$, and the high corrections returned at low
latitudes should be used with caution at this stage. Also, $\epsilon_u$ and
$\epsilon_v$
vary in similar fashion as function of $b$, thus giving further support to
their correlation, and meaning that above $\sim 10^\circ$ from the plane, the
$u-v$ index is affected by $\sim0.06$ mag at most. 

\subsection{Comparison to other methods for zero-points determination}

The method used here to improve photometric zero-points relies on stellar
effective temperatures of a number of stars across the sky. In the literature
there exist similar other methods, at least conceptually, where stellar
properties are used to improve zero-points of large scale photometric surveys.
One rather common technique uses the stellar-locus regression, where the stellar
locus defined by stars in various colour-colour planes is assumed to be
universal, and photometric zero-points in different frames are varied to match
this assumed location \citep[e.g.,][]{macdonald04,ivezic07,covey07,high09}.
Another method is the stellar-colour regression, where stars with reasonably
similar spectroscopic parameters are assumed to
have same colours \citep{yuan15}. The pros and cons of these methods are largely
discussed in the above literature. Very briefly, the stellar-locus regression
strongly relies on the assumption that stellar properties do not vary across
the different populations observed by a large scale survey. Strictly speaking,
this is not true, as stellar age and metallicity gradients are known to exist
across the Galaxy \citep[e.g.][]{boeche14,c16,ciuca18}. Hence, the
stellar-locus regression
is usually not applied to ultraviolet filters, which are intrinsically more
sensitive to variations of stellar properties \citep{high09}. Also, the
stellar-locus regression necessarily correct for extinction, and it produces
discrepant results if the sources of extincion vary significantly across a
field of view. The stellar-colour regression requires instead the
existence of a few photometrically well calibrated fields from which
spectroscopic reference stars are selected in order to determine the intrinsic
colours for a given set of stellar parameters. Stars with spectroscopic stellar
parameters are then needed across the sky, and the reddening values of these
stars must be known. While these limitations are real, they do not impede
stellar-locus and stellar-colour regressions to achieve an internal precision
of order 1 percent or better \citep[e.g.][]{high09,yuan15}.

The photometric standardization currently done in SkyMapper can also be
regarded as a form of stellar-locus regression. In this case, the
locus is defined by the stellar templates used to derive transformations
from APASS to SkyMapper magnitudes (see summary in Section \ref{app:zp}, and
\citealt{wolf18} for full details). As previously discussed, this approach works
remarkably well for SkyMapper optical filters, but not for the $uv$ ones
because of their sensitivity to stellar parameters (a dependence which is not
accounted for in the stellar-locus approach). The method we have developed in
this paper aim to overcome this limitation, by varying photometric zero-points
until reference stellar effective temperatures from the IRFM are reproduced.
The advantage is that the method is differential with respect to stellar
properties and reddening: the same $\logg$, $\feh$ and $E(B-V)$ adopted to
derive reference effective temperatures are used to implement SkyMapper
photometry of the same stars into the IRFM. Also, the IRFM is only mildly
sensitive to the assumed $\logg$ and $\feh$ of stars, and it readily allows to
map known $\teff$ into photometric zero-points. Correct zero-points can thus
be derived if the absolute zero-point of the $\teff$ scale is known. This last 
requirement limits the number of stars across the sky which can be used for
this purpose.

\section{Reddening coefficients and reddening free indexes}\label{sec:red}

A non-negligible amount of foreground dust is present for stars roughly
beyond $\sim 70$pc \citep[e.g.,][]{lallement03}. Since SkyMapper
saturates around $g \sim 10$ (the exact value varying with seeing conditions),
the above distance implies that sources with absolute magnitudes brighter than
$M_g \sim 6$ will suffer from 
extinction. In other words, this affects all stars observed by SkyMapper,
unless we limit ourselves to nearby dwarfs. In this Section we provide
extinction coefficients suitable for late-type stars. Users can adopt those
together with their preferred source of reddening estimates to unredden
observed photometry, before applying the calibrations we provide in Section
\ref{sec:colorte} and \ref{sec:feh}. We also lay out the formalism to use
extinction coefficients to create reddening free pseudo-colours and -magnitudes.
\begin{figure*}
\begin{center}
\includegraphics[scale=0.16]{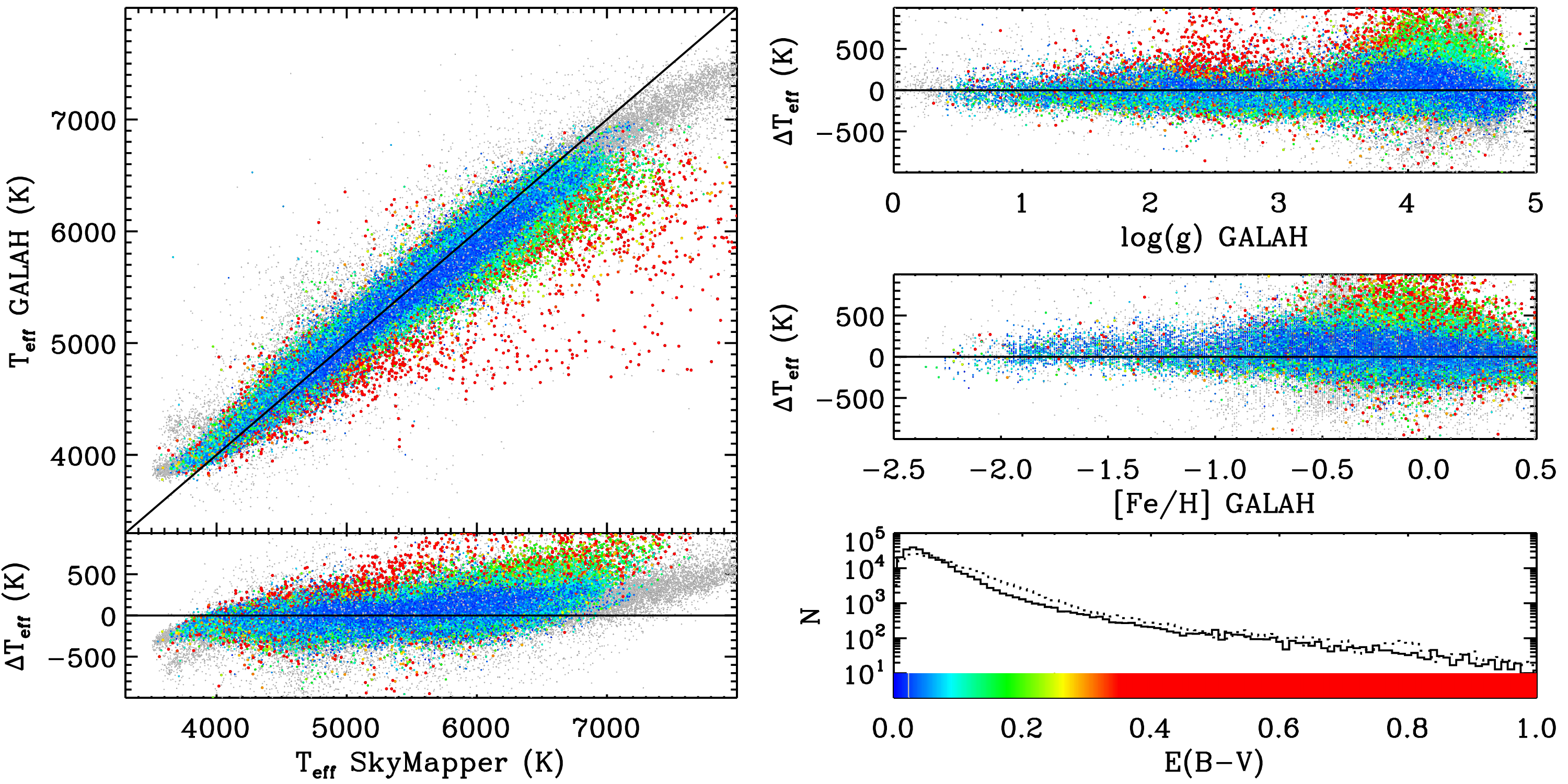}
\caption{Comparison between $\teff$ derived implementing SkyMapper photometry
  into the IRFM, and the GALAH spectroscopic survey \citep{buder18}. Residuals
  (SkyMapper$-$GALAH) are shown as function of stellar parameters, and colour
  coded by reddening according to the scale in the bottom right panel.
  Dotted histogram is reddening from \citet{sfd98}, while continuous histogram
  shows the rescaled values we adopt. Grey points are stars flagged as
  unreliable in GALAH.}
\label{fig:galah}
\end{center}
\end{figure*}

Extinction is usually parametrized as a function of
reddening $E(B-V)$, and $R_V$. The latter is the ratio of total to selective
extinction in the optical, found to be $\simeq 3.1$ for most Galactic
sightlines \citep[e.g.,][]{sms16}. It can be shown that a given
$E(B-V)$ and $R_V$ will affect stars of different spectral types differently
\citep[e.g.,][]{cv14}. For example, the extinction coefficients reported in
\cite{wolf18} are based on a flat spectrum, and the \cite{fitz99} extinction
law. In our implementation of the IRFM, we adopt the \cite{ccm89}/\cite{od94}
extinction law, and iteratively compute extinction coefficients using a
synthetic spectrum at the $\teff, \logg$ and [Fe/H] of each star to deredden
them. In practical terms, extinction coefficients are rather constant, but for
the bluest filters at the coolest $\teff$. Extinction coefficients for the
SkyMapper system are given in Table \ref{tab:ext}. Once extinction
coefficients are known, unreddened magnitudes
$m_{\zeta,0}=m_\zeta-R_\zeta E(B-V)$ and colours
$(\zeta-\eta)_0=(\zeta-\eta)-E(\zeta-\eta)$
$=(\zeta-\eta)-(R_\zeta-R_\eta)E(B-V)$ can be derived. 

Reddening-free pseudo-colours $c_{PS}$ and pseudo-magnitudes $m_{PS}$ can also
be built as follows:
\begin{equation}
c_{PS} = (\zeta-\eta) - X (\xi-\ni)
\end{equation}
where $\zeta,\eta, \xi$ and $\ni$ are any combination of SkyMapper filters, and
$X$ is a multiplicative factor such that any dependence on reddening cancels
out. It can be easily proved that this conditions is met when:
\begin{equation}
X=\frac{R_\zeta-R_\eta}{R_\xi-R_\ni}.
\end{equation}
Similarly, for pseudo-magnitudes:
\begin{equation}
m_{PS} = \zeta - X (\eta-\xi),
\end{equation}
where 
\begin{equation}
X = \frac{R_\zeta}{R_\eta-R_\xi}.
\end{equation}
It must be pointed out that the above reddening-free indices are meaningful
only over the $\teff$ regime where extinction coefficients are nearly constant.
Also, we remark that the use of reddening-free indices is often a trade-off: in
fact while they bypass the dependence on reddening, they correlate more
poorly with stellar parameters. 
\begin{table}
\centering
\caption{Extinction coefficients $R_\zeta$ for a solar 
  temperature star. Notice that for a nominal $E(B-V)$, the excess in any
  given colour combination is $E(\zeta-\eta)=(R_\zeta-R_\eta)E(B-V)$.}\label{tab:ext}
\begin{tabular}{cccccc}
\hline 
 $R_u$ &  $R_v$ &  $R_g$ &  $R_r$ &  $R_i$ &  $R_z$  \\ 
\hline
$4.88$ & $4.55$ & $3.43$ & $2.73$ & $1.99$ & $1.47$ \\
  \hline
\end{tabular}
\begin{minipage}{0.48\textwidth}
  $R_u$ has a strong dependence on $\teff$, which can be fit as
  $R_u=4.95- 2.6\times10^{21}\teff^{-6}$. For $R_g=3.68- 1471\times\teff^{-1}$.
  For the remaining filters, reddening coefficients vary less than $\sim0.1$
  over the range $3500$~{K}$\,< \teff < 10000$~K explored in this work. 
  The values reported here agree with the fit at the solar value from Table B1
  of \citet[][where the fit for $u$ and $g$ band is valid on a much smaller
  $\teff$ range]{cv18}.
\end{minipage}
\end{table}

\section{SkyMapper meets GALAH}\label{sec:smg}

The GALactic Archaeology with HERMES (GALAH) is a stellar spectroscopic survey
conducted on the Anglo-Australian Telescope \citep{galah}. GALAH stellar
parameters are obtained with ``The Cannon'' \citep{cannon}, a data-driven
approach calibrated upon a training set that covers the FGK-type stars
\citep[see][for further details]{buder18}. Currently, nearly half million stars
have been observed
and analyzed, with over $270000$ spectra in common with SkyMapper DR1.1. 
Here, we apply the IRFM on all these stars, and check the performance of
data-driven $\teff$ determination in GALAH DR2 \citep{buder18}.
\begin{figure*}
\begin{center}
\includegraphics[scale=0.2]{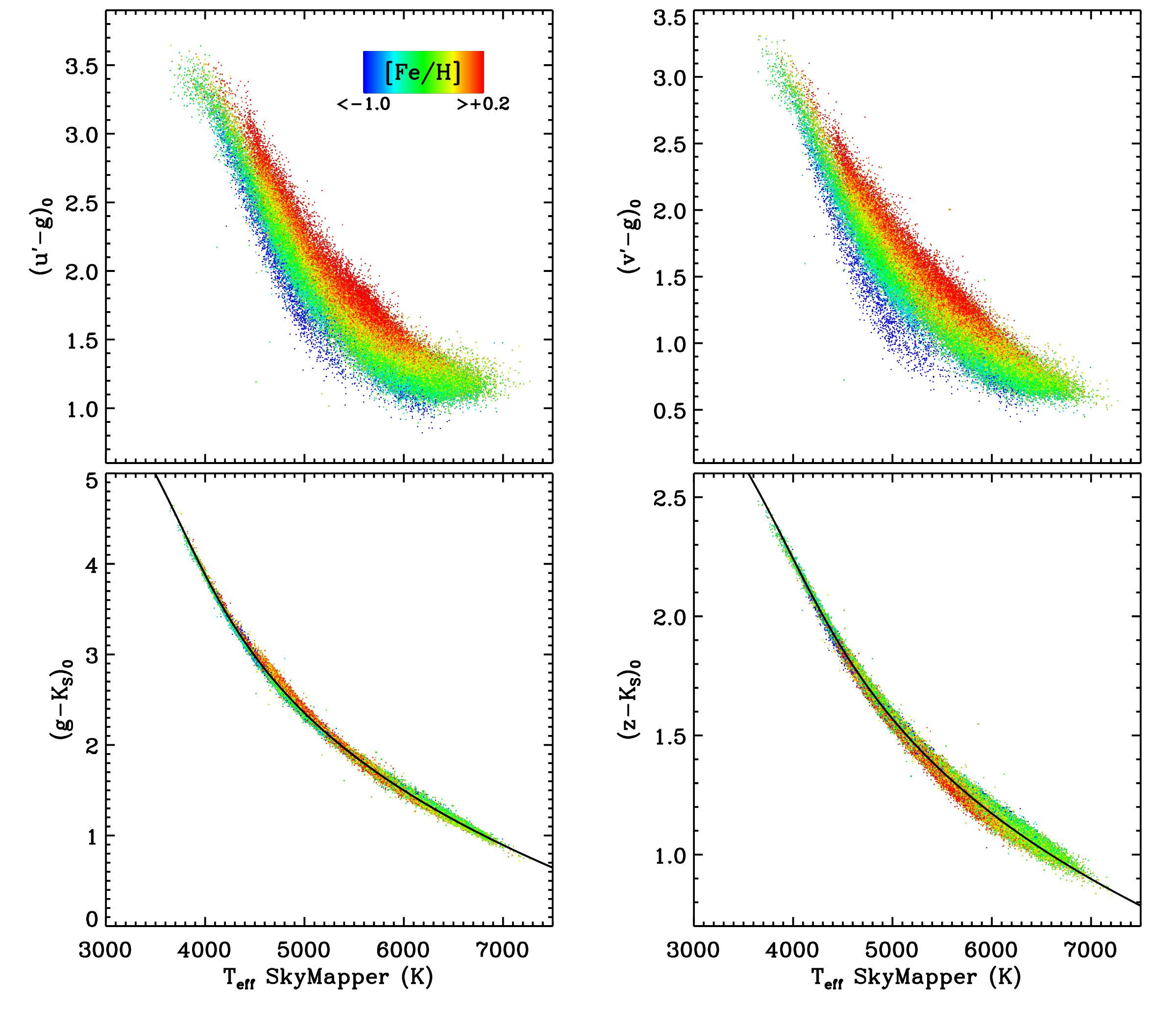}
  \caption{Colour-temperature relations for a few combinations of SkyMapper
    and 2MASS filters. See Section \ref{sec:cal} for the definition of $u'$
    and $v'$. All colours are dereddened using $E(B-V)$ as described
    in the text. Only stars with $E(B-V)<0.05$ are shown. Stars are
    colour-coded according to their GALAH metallicity,
    with the scheme indicated in the inset on the upper left panel.
    For $\feh <-1.0$ dex ($>0.2$ dex) the colour is kept fixed to blue (red).
    Eq.~(\ref{eq:gK}) and (\ref{eq:zK}) are shown as continuous lines in the
    bottom panels.}
\label{fig:colte1}
\end{center}
\end{figure*}

\begin{figure*}
\begin{center}
\includegraphics[scale=0.2]{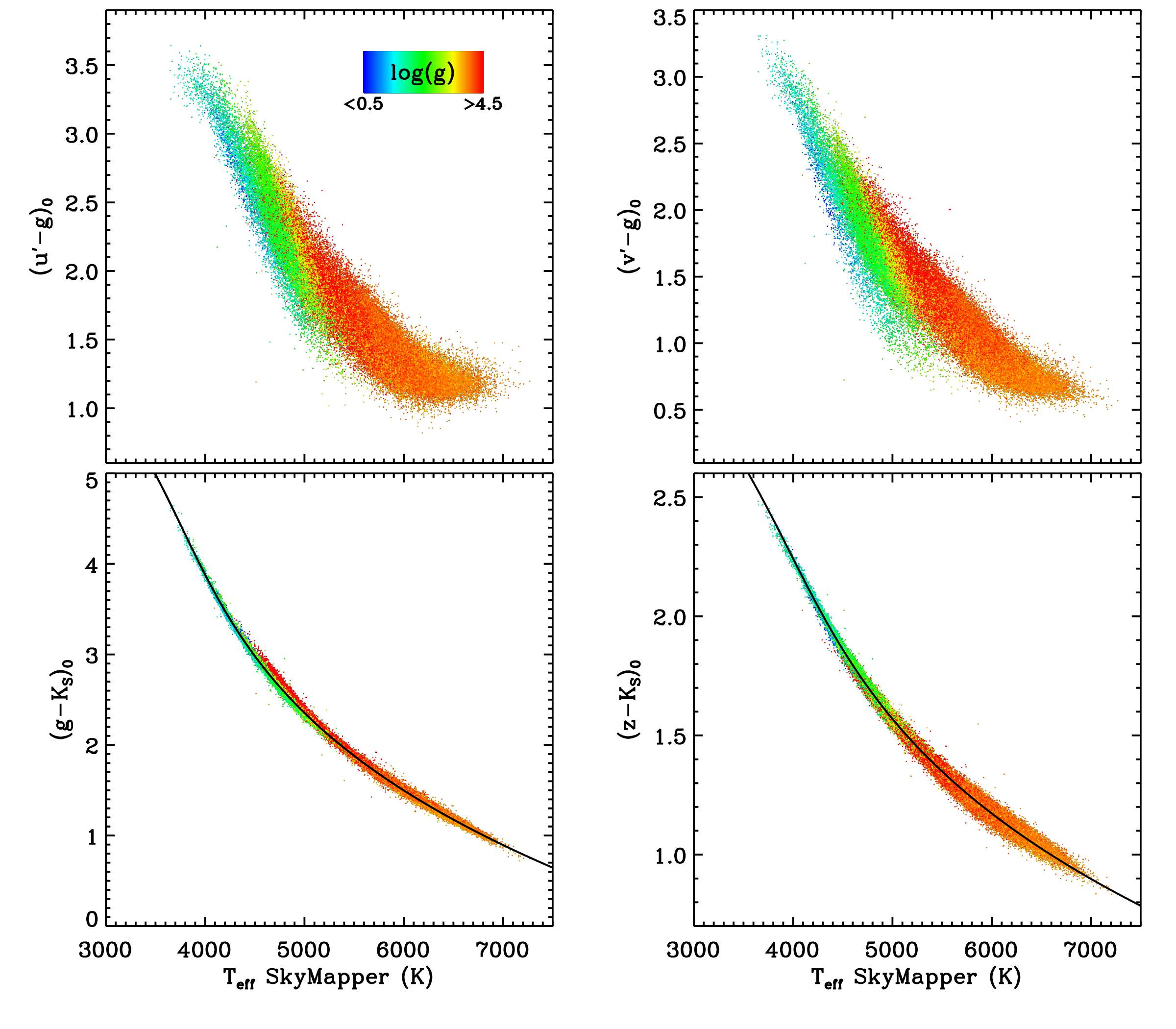}
  \caption{Same as previous figure, but with stars colour-coded according to
    their GALAH surface gravity. For $\logg <0.5$ dex ($>4.5$ dex) the colour
    is kept fixed to blue (red).}
\label{fig:colte2}
\end{center}
\end{figure*}

Apart from a few pointings along the plane, nearly all of the SkyMapper $\cap$
GALAH targets have Galactic latitudes $|b|>10^{\circ}$, meaning that the most
obscured and patchy region of the Galactic plane are avoided. Yet, reddening
can have a non-negligible contribution, and must be taken into account in
photometric methods. For each target we rescale $E(B-V)$ from \cite{sfd98}
using the same procedure developed for RAVE DR5 \citep{kunder}, and which is
solely based on the intrinsic colour of red clump stars \citep[as described in
more details in][]{c14a}. 

We implement the IRFM exploring different combinations of the photometric
zero-points
derived in the previous Section. Because of the zero-point spatial variations
affecting $u$ and $v$ bands, and their small flux contribution (when other
SkyMapper bands are also implemented), we adopt $\teff$ derived using only
$grizJHK_S$ in the IRFM. Notice however that we also derived temperatures
including $uv$ photometry as a check, and verified the effect to be rather
minor. The mean difference and scatter is of order few Kelvin, and $30~$K
respectively, either using the constant zero-points from Table \ref{tab:zp}, or
the spatially dependent ones from Eq.~(\ref{eq:ugb}) and (\ref{eq:vgb}).

Figure \ref{fig:galah} shows the comparison between $\teff$ from the IRFM and
GALAH, colour-coded by the adopted $E(B-V)$. For low reddening regions, the
agreement is usually excellent across the entire stellar parameter range, and
it degrades in regions of high extinction, where $\teff$ from the IRFM are
typically hotter (thus, indicating that in these regions reddening is still
preferentially overestimated, despite our rescaling). Spectra labelled as
unreliable in GALAH (data reduction or Cannon flags $\ne 0$) are plotted in
grey. Effectively all of the stars above $7000$~K are flagged in GALAH, because
of the lack of a training set in this regime, forcing the data-driven approach
to extrapolate the determination of stellar parameters. The IRFM indicates
that the data driven-approach underestimates effective temperatures in this
regime, saturating at $8000$~K which is the limit of the grid of model
atmospheres used for the training set (we checked that this trend is not an
artefact of stars affected by high values of extinction). This comparison
shows how well calibrated effective temperatures from the IRFM can be helpful to
improve spectroscopic pipelines. 

After removing flagged spectra, the SkyMapper$-$GALAH mean (median)
$\Delta\teff$ is $61$~K ($49$~K) with a scatter of $183$~K when stars
are considered irrespectively of their reddening. The above numbers reduce to
$\Delta\teff = 51$~K ($50$~K) with a scatter of $132$~K when restricting to
$E(B-V)<0.10$, and $\Delta\teff=12$~K ($12$~K) with a scatter of $123$~K for
$E(B-V)<0.01$.
This suggests that reddening can easily introduce systematics of order of
a few tens of K on the zero-point of the $\teff$ scale, and it is the primary
source of uncertainty rather than the photometric zero-points when determining
$\teff$.
Further below we explore the sensitivity of SkyMapper colours
to $\feh$ and $\logg$ from GALAH, and $\teff$ from the IRFM.

\subsection{The sensitivity of SkyMapper colours to stellar parameters}\label{sec:cal}

For spectral types ranging from approximately F to early M, we explore the
dependence of SkyMapper colours on stellar parameters. For the latter we adopt
$\feh$ and $\logg$ from the GALAH sample (using only non-flagged stars),
whereas effective temperatures come from the IRFM.

In all instances, colours have been dereddened with the $E(B-V)$ derived in
Section \ref{sec:smg}, and using extinction coefficients appropriate to the
$\teff, \logg$ and $\feh$ of each star (as discussed in Section \ref{sec:red}).
In the remainder of the paper, all plots and calibrations are corrected for
reddening, and this is indicated by the $0$ subscript. Users
should always correct for reddening their input photometry before applying our
calibrations. Concerning the zero-points offsets discussed in Section
\ref{sec:irfm} and \ref{app:zp}, a few remarks are necessary. Constant
zero-point offsets are of no importance when a calibration between observed
colours and stellar parameters is built, since any zero-point is automatically
factored into the calibration. Hence, zero-point corrections must not be applied
to Eq.~(\ref{eq:gK}) and (\ref{eq:zK}). On the contrary, spatially dependent
zero-points must be corrected for, and this is the case for the metallicity
calibration discussed further below. In the rest of the paper, we define 
$u'=u-\epsilon_u$ and $v'=v-\epsilon_v$, where $\epsilon_u$ and $\epsilon_v$
are given in Eq.~(\ref{eq:vgb}).

\subsubsection{Colour-$\teff$ relations}\label{sec:colorte}

$\teff$ is the stellar parameter to which colours are most sensitive, and
arguably the most needed e.g., to constrain spectroscopic analyses. 
Figure \ref{fig:colte1} and \ref{fig:colte2} show the colour-$\teff$ relations
derived from the IRFM in a selected number of colour indices, to highlight their
dependence on metallicity and
surface gravity. SkyMapper photometry performs satisfactorily to separate stars
with different stellar parameters, in particular when using the $u$ and $v$
bands. From these figures, the interplay between metallicity and surface
gravity in driving changes in photometric colours is obvious, besides sample
selection effects (e.g., at the
coolest $\teff$ essentially all of the stars are giants, since M dwarfs are
not analysed in GALAH). This means that is not straightforward to provide a
unique functional form that works for all colour indices, and accounts at the
same time for $\logg$ and $\feh$ effects. At the same time, in practical
instances
users are often interested to estimate $\teff$ without prior knowledge of
the metallicity and surface gravity of stars. We find that the $(g-K_S)_0$ and
$(z-K_S)_0$ show a tight correlation with $\teff$, and little sensitivity to
$\feh$ and $\logg$. For these colours, third-order polynomials suffice
to fit the data well:
\begin{displaymath}
\teff= 9056.01 - 2732.89\,(g-K_S)_0\;+ 
\end{displaymath}
\begin{equation}\label{eq:gK}
\phantom{\teff=}522.40\,(g-K_S)_0^2 - 39.66\,(g-K_S)_0^3
\end{equation}
which has $\sigma=33$~K and is valid for $0.46<(g-K_S)_0<4.65$, and 
\begin{displaymath}
\teff= 12884.70 - 9336.50\,(z-K_S)_0\;+ 
\end{displaymath}
\begin{equation}\label{eq:zK}
\phantom{\teff=}3567.35\,(z-K_S)_0^2 - 522.11\,(z-K_S)_0^3
\end{equation}
which has $\sigma=59$~K and is valid for $0.68<(z-K_S)_0<2.48$.

\subsubsection{Surface gravity}

\begin{figure}
\begin{center}
\includegraphics[scale=0.15]{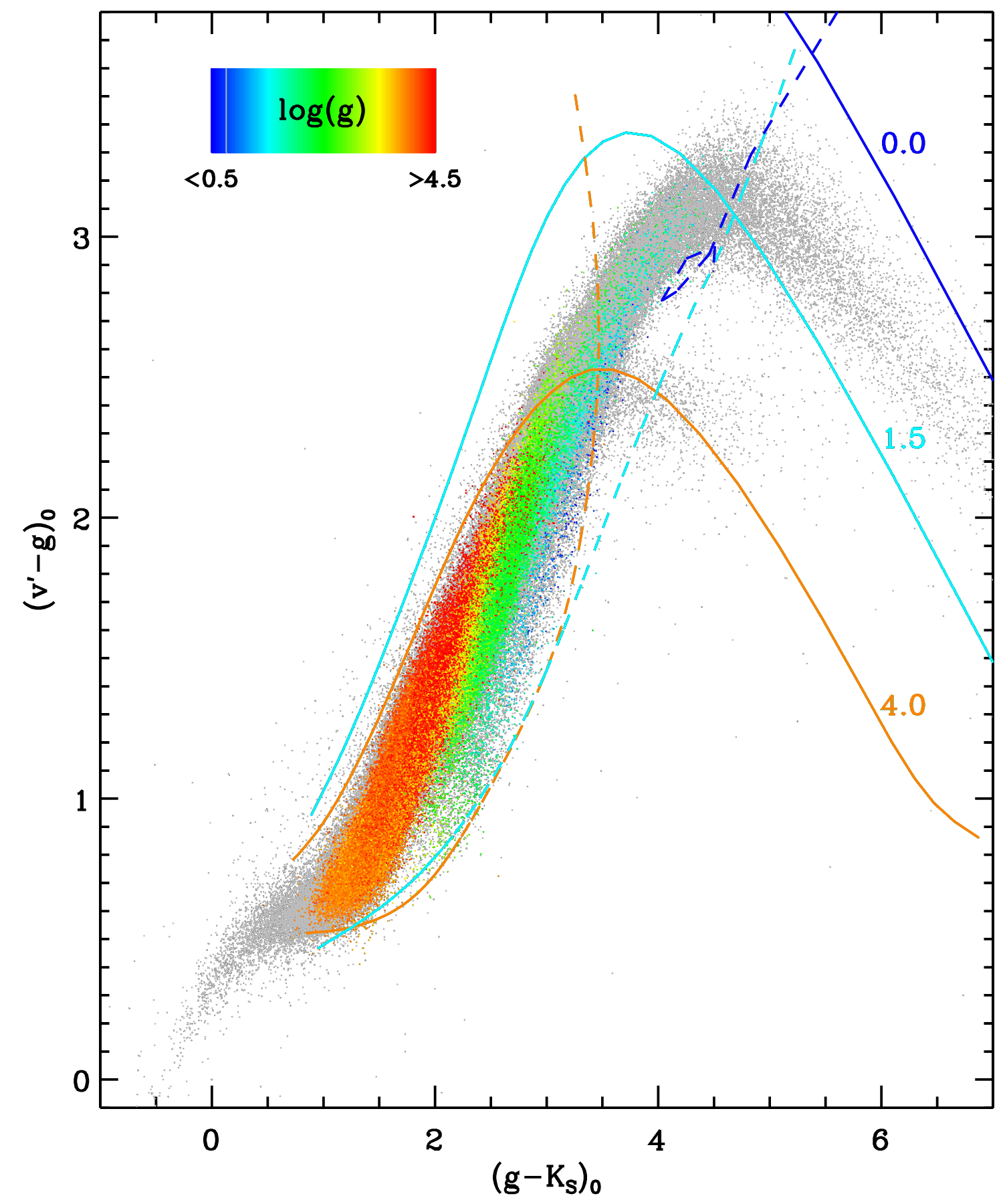}
\caption{Colour-colour plane with GALAH stars coded by their $\logg$
  as per inset panel. Grey dots are stars from RAVE DR5. Continuous lines are
  synthetic colours from \citet{cv14,cv18} at the $\logg$ values indicated.
  Continuous and dotted lines are for $\feh=0.5$ and $-4.0$, respectively.}
\label{fig:gkvg1}
\end{center}
\end{figure}

The gravity sensitivity of SkyMapper filters has already been explored in some
detail in \citet[][see their figure 16 for examples of colour-colour plots
discriminating dwarfs and giants]{wolf18}. Here, we do not repeat that exercise,
but rather focus on the gravity sensitivity of the $(v'-g)_0$ vs. $(g-K_S)_0$
colours, which we will use as $\feh$ indicators in Section \ref{sec:feh}. Figure
\ref{fig:gkvg1} shows the the dependence of this colour combination on
$\logg$. We have already discussed how GALAH does not derive parameters for very
cool dwarfs. Thus, in addition to the GALAH sample (colour-coded), we also
include stars from RAVE DR5 \citep[][here shown in grey]{kunder}, which has a
larger number of late-type dwarfs and giants. For $(g-K_S)_0 \gtrsim 3.5$ (which
corresponds to $\teff \lesssim 4200$~K), dwarf and giant stars clearly define
distinct sequences in this colour plane. However, at bluer colours, there is
very little dependence on $\logg$, and this is qualitatively confirmed by synthetic
stellar colours \citep[see the discussion in][for the performance and
  limitation of stellar synthetic colours, in particular at blue wavelengths,
  and for cool stars]{cv14,cv18}. 
\begin{figure*}
\begin{center}
\includegraphics[scale=0.17]{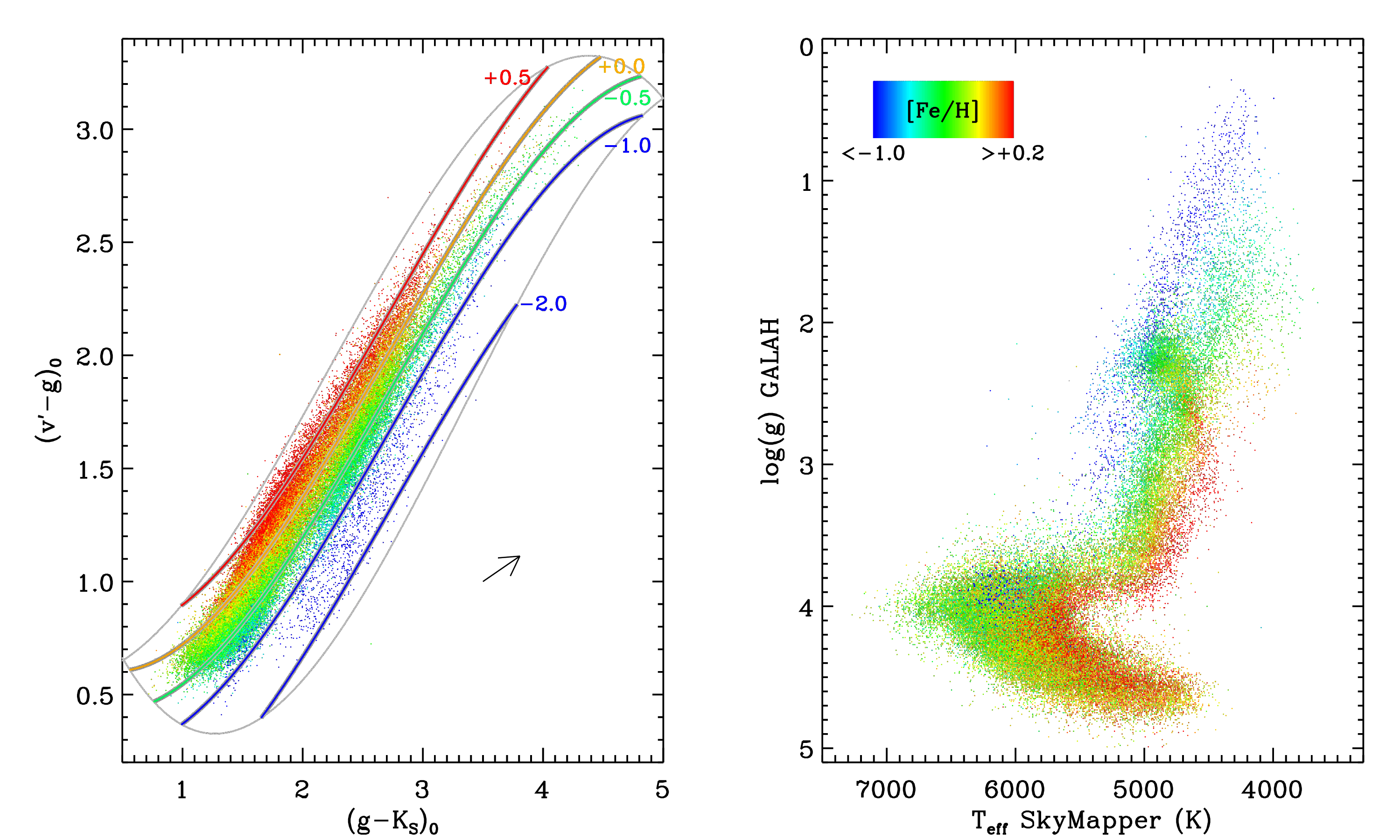}
\caption{Left-panel: colour-colour plane with GALAH stars coded
  by their $\feh$ as per inset panel on the right. Grey lines define the
  boundary of our metallicity calibration, while continuous coloured lines trace
  Eq.~(\ref{eq:mecal}) at indicated values
  of $\feh$. The arrow shows the direction of the reddening vector with length
  corresponding to $E(B-V)=0.1$. Right-panel: Kiel diagram for the
  same stars. In both panels, only stars with $\ebv < 0.05$ and
  $|b|>20^\circ$ are shown, although relaxing these conditions does not
  qualitatively change the plots.}
\label{fig:gkvg2}
\end{center}
\end{figure*}

\subsubsection{Colour-$\feh$ relation}\label{sec:feh}

The determination of photometric metallicities is one of the goals behind the
design of SkyMapper filters, in particular the $v$ band. In order to explore
the correlation of different colour indices with stellar parameters we use
Principal Component Analysis \citep[PCA, e.g.,][]{fw99}. Depending on the colour
combination, we see the clear presence of up to three principal components.
Regardless of the colour index though, the first component always correlates
strongly with $\teff$, while the second and the third correlate with different
strength to $\feh$ and $\logg$, respectively. PCA thus confirms what we already
knew, i.e., the colours of stars depend primarily on their effective
temperature, while metallicity and surface gravity are less important, but
non-negligible in certain bands (in fact, see Figure \ref{fig:colte1} and
\ref{fig:colte2}). However, the presence of a correlation does not guarantee
that a useful calibration between stellar parameters and colour indices can
always be found. For example, while all indices involving the $u$ band
correlate with $\logg$, no calibration can be found beyond a qualitative
separation between late-type dwarfs and giants. However, in the $(v'-g)_0$ vs.
$(g-K_S)_0$ colour plane we find a strong correlation between the second
principal component and $\feh$. We derive the following calibration between
colours and metallicities:
\begin{displaymath}
\textrm{[Fe/H]} = \frac{-0.1815+0.1848\,(v'-g)_0-0.1630\,(g-K_S)_0}{0.0649} + 
\end{displaymath}
\begin{equation}\label{eq:mecal}
0.8501+3.6086\,(g-K_S)_0 - 1.3735\,(g-K_S)_0^2+0.1684\,(g-K_S)_0^3
\end{equation}
where the first term is derived from PCA analysis, and the third order
polynomial in $(g-K_S)_0$ is obtained fitting the residual as function of this
colour index. We explored the use of more colour terms, as well as higher order
polynomials, but found that our metallicity calibration did not improve.
We suspect that this might be due to photometric uncertainties, where the gain
of using more colours trades off with an increased error budget. Thus, we
decide
to adopt this rather minimalistic functional form, which also has the advantage
of being broadly parallel to the reddening vector
(see Figure \ref{fig:gkvg2}). Our calibration is derived using only stars with
$E(B-V)<0.05$, and located at Galactic latitudes $|b|>20^\circ$ to avoid
introducing strong dependencies on zero-point corrections. We define two
fiducial lines beyond which our calibration should not be extrapolated
$P_1<(v'-g)_0<P_2$ (grey lines in left-panel of Figure \ref{eq:mecal}), where:
\begin{equation}
P_1=1.3067-1.6731(g-K_S)_0 + 0.8129(g-K_S)_0^2 - 0.0810(g-K_S)_0^3
\end{equation}
and
\begin{equation}
P_2=0.5783-0.0719(g-K_S)_0 + 0.4624(g-K_S)_0^2 - 0.0691(g-K_S)_0^3.
\end{equation}
With these criteria, our training sample comprises over $70,000$ stars, and
the standard deviation of our metallicity calibration is $0.21$~dex. Also,
these fiducials limit the metallicity range of our calibration, which applies
down to $\feh \simeq -2$. We have verified that extending our calibration to
more metal poor stars leads to mixed results. There are several reasons for
this: while model atmosphere fluxes and isochrones indicate that Skymapper
$v-g$ related colors should be useful down to metallicities of $-4$, current
photometric errors in SkyMapper ultraviolet bands (see discussion in Section
\ref{app:zp}) prevent to exploit its full potential to reliably single out the
most metal poor stars. In addition, because of the increasing fraction of
carbon-enhanced stars below $-2$ \citep[e.g.,][]{yong13}, the B-X band of the
CH molecules dump the flux around the location of the $v$ band\footnote{We
  remark however that the CH G-band (A-X) falls within the $g$ filter and does
  not contaminate the Skymapper $v$ band.}, hence
mimicking a higher metal content. For a detailed investigation of the
performances of SkyMapper photometry to identify extremely metal-poor stars,
where other filter combinations are more appropriate, we refer to
\cite{dacosta}.

\begin{figure*}
\begin{center}
\includegraphics[scale=0.15]{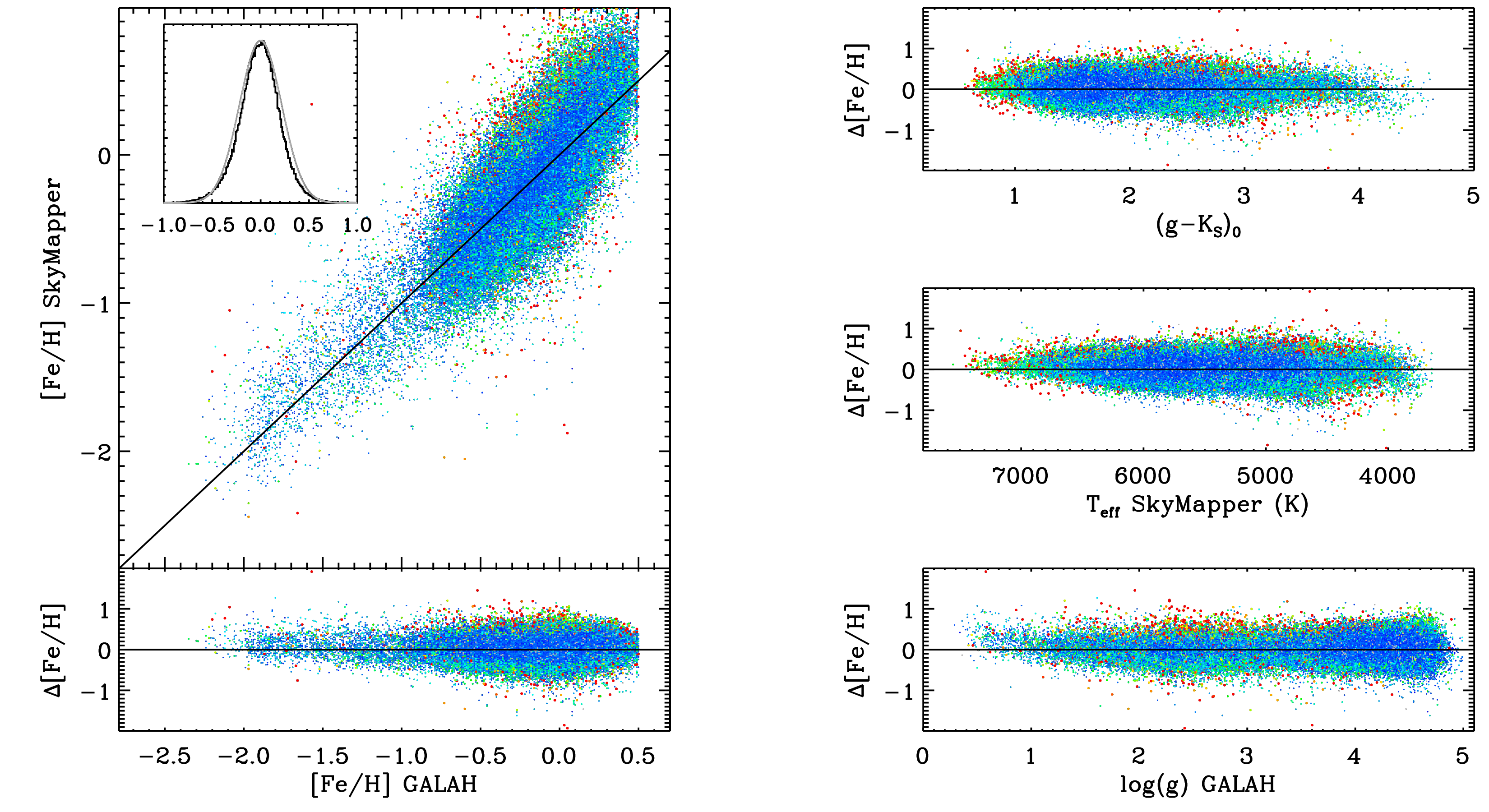}
\caption{Left-panels: GALAH versus SkyMapper photometric metallicities
  for over $160,000$ stars using our calibration (top), and residuals
  (SkyMapper$-$GALAH) fitted with a Gaussian of width $0.22$~dex overplot
  (inset),
  and as function of $\feh$ (bottom). Right-panels: residuals as function of
  $(g-K_S)_0$, $\teff$ and $\logg$. In all figures, stars are colour-coded
  according to their reddening, as per the palette in lower-right panel of 
  Figure \ref{fig:galah}.}
\label{fig:fehcal}
\end{center}
\end{figure*}

Figure \ref{fig:fehcal} compares the metallicities derived from
Eq.~(\ref{eq:mecal}) against the entire GALAH sample, irrespective of
reddening and Galactic latitude, thus comprising over
$160,000$ stars. The standard deviation is virtually unchanged, $0.22$~dex,
confirming that reddening has a minimal impact upon our calibration. Also,
the fact that we now probe latitudes closer to the plane, and still obtain
satisfactory metallicities speaks well of our zero-point corrections. 
The most
discrepant points in Figure \ref{fig:fehcal} are indeed those with the highest
reddening, but a large value of reddening does not univocally imply that
photometric metallicities are unreliable. There is a large number of stars at
high $E(B-V)$, for which spectroscopic and photometric metallicities are
in good agreement (although in Figure \ref{fig:fehcal} they are hidden behind
an overwhelming number of stars at low reddening). 
Taking into account that GALAH metallicities are precise to within $0.1$~dex,
this gives us confidence that photometric metallicities can be derived to
a precision of $0.2$~dex from our calibration. Residuals as function
of colour, $\teff$ and $\logg$ show that photometric metallicities are good
across the entire parameter space explored, with increasing scatter and a mild
offset towards the highest and lowest gravities, respectively. Although
our metallicity calibration works well for both dwarfs and giants over the
parameter space explored, we remark that GALAH does not provide parameters for
dwarfs with $\teff \lesssim 4500$~K (see Kiel diagram in Figure \ref{fig:gkvg2}), where we expect
to see a bifurcation between the dwarf and the giant sequence (Figure
\ref{fig:gkvg1}). Thus, the decreasing scatter in the residuals towards the
reddest colours (and coolest $\teff$) carries a sample selection effects: it
reflects the adequacy of the calibration for cool giants, but it does not
warrant its use for cool dwarfs. The fiducial $P_1$ is intended to limit
contamination from cool dwarfs, although it does not remove them entirely.
Thus, we advise using {\it Gaia} parallaxes to exclude cool dwarfs (see next
Section), as well as to preferentially apply our metallicity calibration for
$(g-K_S)_0\lesssim3.5$, where the effect of surface gravity on the metallicity
calibration is minor. 

\begin{figure*}
\begin{center}
\includegraphics[scale=0.15]{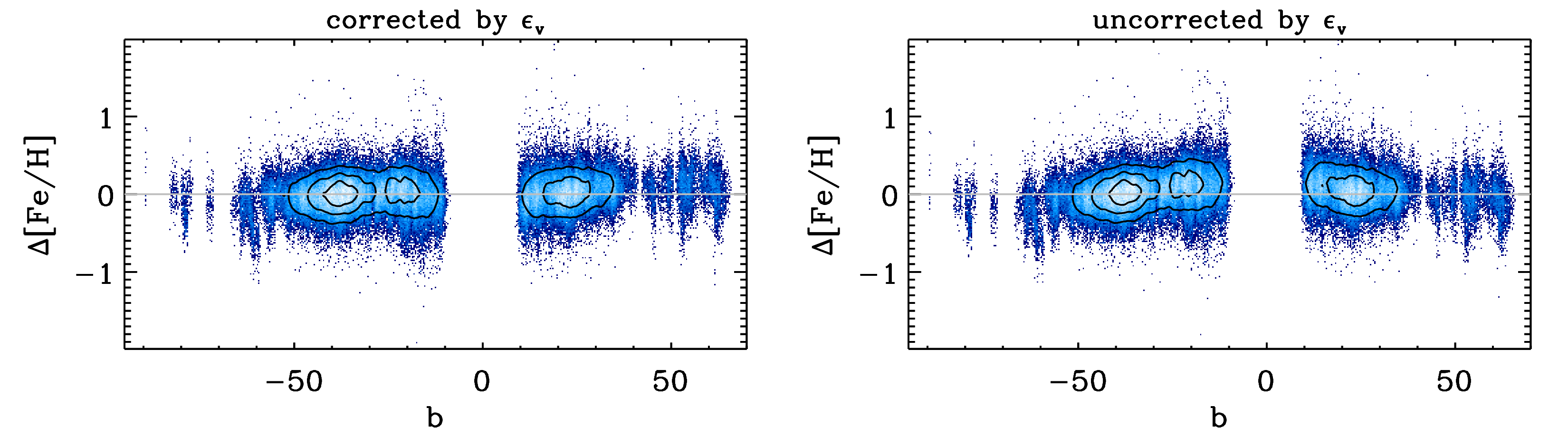}
\caption{Left-panel: metallicity residuals (spectroscopic minus photometric)
  as function of Galactic latitude when the calibration is derived using $v'$.
  Right-panel: same as left panel, but using instead $v$ to derive the
  metallicity calibration. Colours indicate the density of stars, from
  highest (light-blue) to lowest (dark-blue). Contour levels are also shown to
  make the trend more clear.}
\label{fig:gboff}
\end{center}
\end{figure*}

Finally, Figure \ref{fig:gboff} shows the residual of photometric versus
spectroscopic metallicities as function of Galactic latitude. 
No trend is seen when $v'$ magnitudes are used for the metallicity calibration,
whereas this is
not the case for $v$: a clear trend appears as function of $b$, and this could
e.g., lead to biases when measuring
vertical metallicity gradients. We remark that the metallicity calibration is
obtained using only stars at $|b|>20^\circ$, as well as with an entirely
different sample (and method) than the one used to study the spatial variation
of zero-points (Section \ref{app:zp}). The fact that a trend as function of $b$
is now seen in Figure \ref{fig:gboff} when we do not correct $v$ magnitudes,
gives us further confidence that the zero-point variations we uncover are
real.  

\section{External validations \& comparison to Sloan}

To further check the performance of our metallicity calibration, we first
compare our photometric $\feh$ with two spectroscopic surveys other than GALAH,
and then use SkyMapper+2MASS photometry to derive a metallicity map of the
Milky Way. In all instances, we correct for reddening with the same
prescription of Section \ref{sec:smg}. Finally we discuss the sensitivity to
metallicity of the SkyMapper $v$ filter in comparison to the Sloan Digital Sky
Survey (SDSS) $u$ band, the first survey to provide ultraviolet photometry for
several million sources across the sky \citep{ivezic07}.

\begin{figure*}
\begin{center}
\includegraphics[scale=0.15]{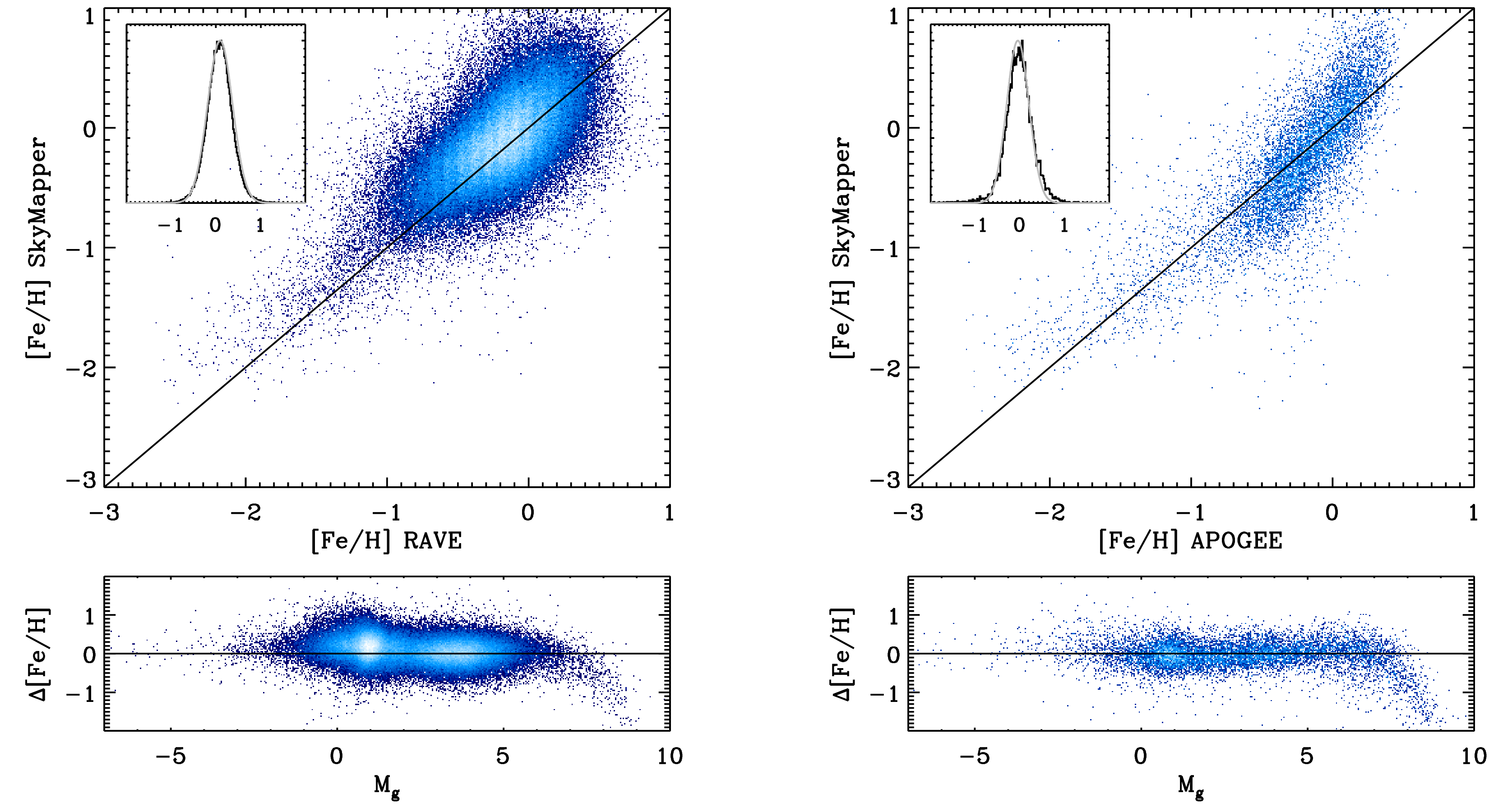}
\caption{Left-panel: RAVE DR5 versus SkyMapper metallicities for over $140,000$
  stars using our calibration (top), and residuals fitted with a Gaussian of
  width $0.28$ dex (inset). Only stars with $M_g < 7$, RAVE flags
  {\tt ALGO\_CONV=0, c1=c2=c3=n} and $P_1<(v'-g)_0<P_2$ are used for this
  comparison. Residuals as function of $M_g$ (bottom) are also shown, to
  highlight how the metallicity calibration degrade for $M_g > 7$.
  Right-panel: same as left-panel, but using over $9,500$ stars having ASCAP
  parameters from APOGEE DR14. Only objects without bad flags are used.
  Residuals are fitted with a Gaussian of width $0.25$~dex. Colours indicate
  the density of stars.}
\label{fig:surveys}
\end{center}
\end{figure*}

The left-hand panel of Figure
\ref{fig:surveys} compares our photometric metallicities against those in RAVE
DR5 \citep{kunder}. There is a mean offset of $0.09$~dex (SkyMapper minus RAVE)
and a scatter of $0.28$~dex, which is consistent with the lower precision of
$\feh$ in RAVE. The right-hand panel compares our metallicities against those in
APOGEE DR14 \citep{dr14}. In this case there is a smaller offset of
$-0.01$~dex (SkyMapper
minus APOGEE) and scatter of $0.25$~dex. The advantage of these comparisons is
the presence of cool dwarfs which are not part of the GALAH sample. Using
Gaia's parallaxes \citep{Gaia18}, we clearly see that metallicity
residuals deteriorate for  $M_g \geqslant 7$, which we adopt as the absolute
magnitude limit beyond which our metallicity calibration should not be used. 

\begin{figure*}
\begin{center}
\includegraphics[scale=0.4]{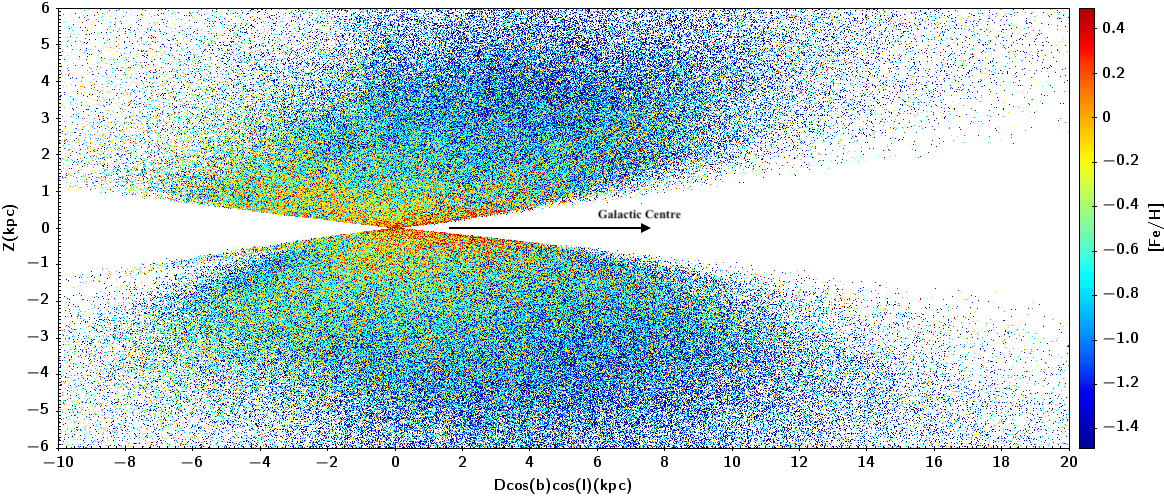}
\caption{Milky Way metallicity map using $\simeq 9$ millions stars for which our
  calibration can be applied, and with Gaia DR2 parallaxes. In this Cartesian
  frame, the Sun is located at $(0,0)$, where $Z$ is height from the plane,
  $D$ is the distance and $(l,b)$ are Galactic coordinates. The direction to
  the Galactic Centre (approximately at 8 kpc) is also indicated.}
\label{fig:GA}
\end{center}
\end{figure*}

Finally, Figure \ref{fig:GA} shows a metallicity map of the Milky Way derived
using $\simeq 9$ million stars with Gaia parallaxes, good SkyMapper and 2MASS
$vgK_S$ photometry and applying Eq. \ref{eq:mecal} within its range of validity.
For the sake of this plot we do
not apply any requirement on the quality of parallaxes, since the goal is
mostly illustrative. We verified though, that restricting to parallaxes better
than 10 percent and adopting the quality cuts in \cite{arenou18} we still see
the same metallicity trends, although with a much reduced number of stars, and
probing a smaller volume. The empty regions close to the plane are areas
currently not targeted by SkyMapper. Nevertheless, we can clearly see high
metallicity stars being preferentially restricted to the Galactic plane, and
the mean metallicity decreasing when moving to higher Galactic height $|Z|$,
transitioning from the thin to the thick disc into the halo, just as expected
from
our knowledge of the  Galaxy. While a proper study of the metallicity
structure would require accounting for target selection effects, and we defer
this to a future
investigation, Figure \ref{fig:GA} gives an example of the kind of studies
SkyMapper photometry will enable. 

\begin{figure*}
\begin{center}
\includegraphics[scale=0.48]{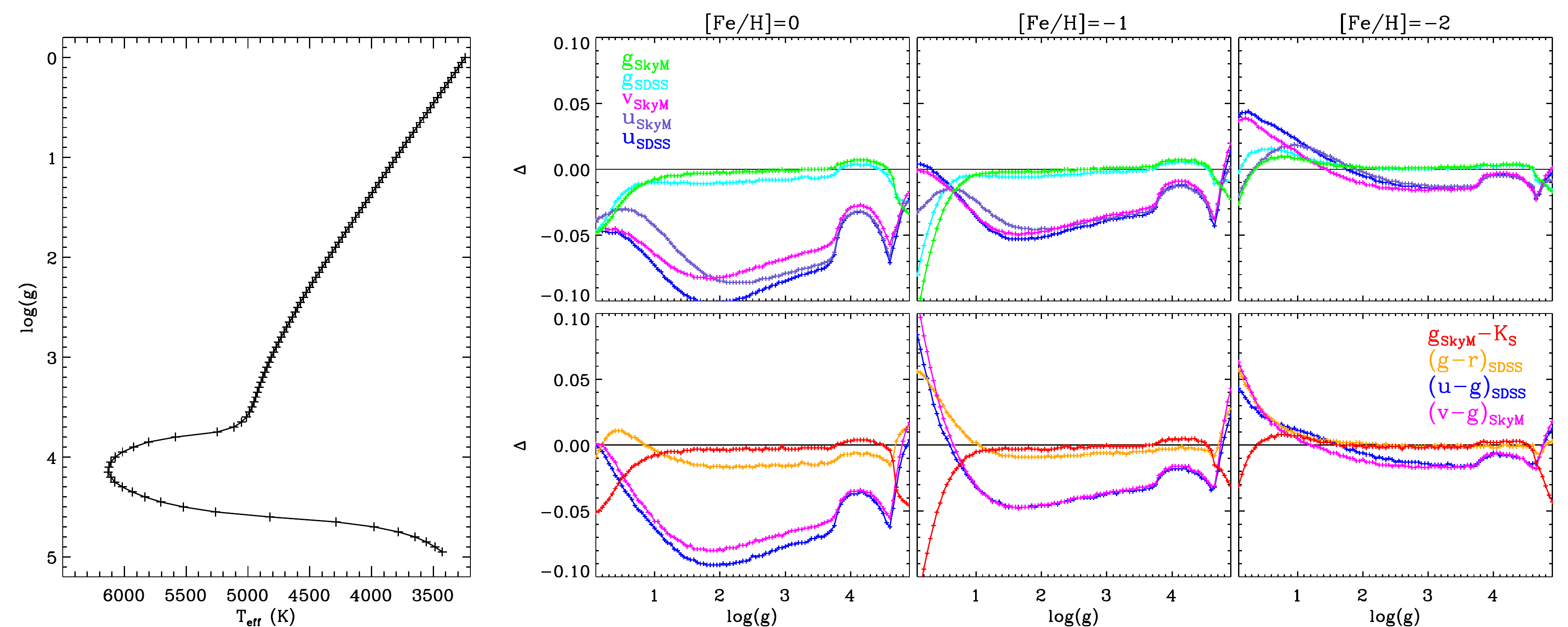}
\caption{Predicted sensitivity to metallicity in a number of filters and
  colour indices using synthetic photometry from \citealt{cv14,cv18}.
  Left-panel: crosses mark the location in the $\teff-\logg$ plane where the
  photometric sensitivity to a change in metallicity is quantified.
  Upper-right panels: change in magnitude ($\Delta$) for Sloan $u,g$ and
  SkyMapper $u,v$ and $g$ filters when metallicity at $0, -1$ and $-2$ (as
  indicated) is decreased by $0.1$~dex. Lower-right panels: same as above, but
  for the colour indices used in our (red and magenta) and Ivezic's
  metallicity calibration (orange and blue).}
\label{fig:pc}
\end{center}
\end{figure*}

Figure \ref{fig:GA} is reminescent of the Milky Way metallicity tomography
done by \cite{ivezic08}, using 2.5 million stars with SDSS colours and
photometric distances. The rms scatter of our metallicity residual
($0.21$~dex) is also similar to their ($0.24$~dex), which is not entirely
surprising since both works crucially rely on the use of one ultraviolet filter:
SkyMapper $v$ for us (centred at $\sim 3800\AA$ with a bandwidth of
$\sim 320\AA$) versus Sloan $u$ for them (centred at $\sim 3600 \AA$ with a
bandwidth of $\sim 540\AA$). However, the functional form we use for
our metallicity calibration has about half the colour terms compared to
\citet[][their eq 4]{ivezic08}.
The theoretical sensitivity of some SkyMapper and
Sloan filters to metallicity is quantified in Figure \ref{fig:pc}, which shows
the change of synthetic magnitudes (upper panels) and colours (lower panels) for
a $0.1$ dex decrease in metallicity at a given $\teff$, $\logg$ and $\feh$. 
The sequence of $\teff$ and $\logg$ sampled (left panel) is typical of the
parameter space covered by late-type stars. Keeping in mind the performances of
theoretical colours in matching real data, our goal here is to single out the
effect on photometry of changing metallicity at given $\teff$ and $\logg$. Thus
we
favour this approach over the use of isochrones, where a change of
metallicity would move isochrones in the $\teff-\logg$ plane as well.
Figure \ref{fig:pc} indicates that the use of the SkyMapper $v$ band yields a
metallicity sensitivity similar to the Sloan $u$, at least over the metallicity
range covered by our calibration. However, the larger bandwidth of Sloan $u$
makes it sensitive to $\logg$, whereas this is not the case for SkyMapper $v$
within the limits we previously discussed.

\section{Conclusions}

In this paper we have conducted a thorough study of SkyMapper DR1.1 photometry.
First, we have checked its standardization; ideally to do so a large number of
absolute flux standards would be needed. Given their
current absence, we have devised a new method based on the effective
temperature of a sample of reference stars to determine photometric
zero-points across the sky. This approach is applicable to any photometric
survey, but it is particularly relevant for SkyMapper, since its zero-points
are not tied to spectrophotometric standard stars, but are obtained from
predicted SkyMapper magnitudes of an ensemble of stars with photometry
from other surveys. The approach currently adopted by SkyMapper works
remarkably well for $griz$, but has limitations in the $uv$ bands. With our
method we have recovered an offset of the $uv$ zero-points that varies as 
a function of Galactic latitude. This variation is expected as a result of the
reddening corrections currently employed in predicting SkyMapper $uv$
magnitudes from external photometry at longer wavelengths.

With a good control over photometric zero-points, we have then applied the
InfraRed Flux Method to derive effective temperatures for all stars in the
GALAH spectroscopic
survey, and provide empirical colour$-\teff$ relations. We have also used
the GALAH spectroscopic metallicities to derive a relation between them and
SkyMapper $v$, $g$, and 2MASS $K_S$ magnitudes.
Our calibrations is validated down to approximately $\feh=-2$, and applies
to late-type giants, and dwarfs with $M_g<7$. The reliability of our photometric
metallicities is further checked against RAVE DR5 and APOGEE DR14, confirming
an overall precision of $0.2$~dex. Finally, using $\sim 9$ million stars with
Gaia parallaxes, we have produced a metallicity map in which we can clearly
trace the mean metallicity decreasing as we move from the thin disc to the thick
disc and then on into the halo, in agreement with what is expected from our knowledge of the
Milky Way's structure. 

\section*{Acknowledgements}
We thank the referee for constructive comments that strengthen the presentation. LC, ADM and DY acknowledge support from the Australian Research Council (grants FT160100402, FT160100206 and FT140100554). Parts of this research were conducted by the Australian Research Council Centre of Excellence for All Sky Astrophysics in 3 Dimensions (ASTRO 3D), through project number CE170100013. This research made use of TOPCAP \citep{topcat}. The national facility capability for SkyMapper has been funded through ARC LIEF grant LE130100104 from the Australian Research Council, awarded to the University of Sydney, the Australian National University, Swinburne University of Technology, the University of Queensland, the University of Western Australia, the University of Melbourne, Curtin University of Technology, Monash University and the Australian Astronomical Observatory. SkyMapper is owned and operated by The Australian National University's Research School of Astronomy and Astrophysics. The survey data were processed and provided by the SkyMapper Team at ANU. The SkyMapper node of the All-Sky Virtual Observatory (ASVO) is hosted at the National Computational Infrastructure (NCI). Development and support the SkyMapper node of the ASVO has been funded in part by Astronomy Australia Limited (AAL) and the Australian Government through the Commonwealth's Education Investment Fund (EIF) and National Collaborative Research Infrastructure Strategy (NCRIS), particularly the National eResearch Collaboration Tools and Resources (NeCTAR) and the Australian National Data Service Projects (ANDS). This work has made use of data from the European Space Agency (ESA) mission {\it Gaia} (\url{https://www.cosmos.esa.int/gaia}), processed by the {\it Gaia} Data Processing and Analysis Consortium (DPAC, \url{https://www.cosmos.esa.int/web/gaia/dpac/consortium}). Funding for the DPAC has been provided by national institutions, in particular the institutions participating in the {\it Gaia} Multilateral Agreement. This publication makes use of data products from the Two Micron All Sky Survey, which is a joint project of the University of Massachusetts and the Infrared Processing and Analysis Center/California Institute of Technology, funded by the National Aeronautics and Space Administration and the National Science Foundation. Funding for RAVE (\url{www.rave-survey.org}) has been provided by institutions of the RAVE participants and by their national funding agencies. Funding for the Sloan Digital Sky Survey IV has been provided by the Alfred P. Sloan Foundation, the U.S. Department of Energy Office of Science, and the Participating Institutions. SDSS-IV acknowledges support and resources from the Center for High-Performance Computing at the University of Utah. The SDSS web site is \url{www.sdss.org}. SDSS-IV is managed by the Astrophysical Research Consortium for the  Participating Institutions of the SDSS Collaboration including the  Brazilian Participation Group, the Carnegie Institution for Science, Carnegie Mellon University, the Chilean Participation Group, the French Participation Group, Harvard-Smithsonian Center for Astrophysics, Instituto de Astrof\'isica de Canarias, The Johns Hopkins University, Kavli Institute for the Physics and Mathematics of the Universe (IPMU) / University of Tokyo, the Korean Participation Group, Lawrence Berkeley National Laboratory, Leibniz Institut f\"ur Astrophysik Potsdam (AIP), Max-Planck-Institut f\"ur Astronomie (MPIA Heidelberg), Max-Planck-Institut f\"ur Astrophysik (MPA Garching), Max-Planck-Institut f\"ur Extraterrestrische Physik (MPE), National Astronomical Observatories of China, New Mexico State University, New York University, University of Notre Dame, Observat\'ario Nacional / MCTI, The Ohio State University, Pennsylvania State University, Shanghai Astronomical Observatory, United Kingdom Participation Group, Universidad Nacional Aut\'onoma de M\'exico, University of Arizona, University of Colorado Boulder, University of Oxford, University of Portsmouth, University of Utah, University of Virginia, University of Washington, University of Wisconsin, Vanderbilt University, and Yale University.

\appendix

\section[]{Monochromatic and in-band fluxes}

With the photon-counting formalist adopted in this paper, the monochromatic
flux associated to the effective wavelength of each SkyMapper magnitude can be
determined from Eq.~(\ref{eq:AB})--(\ref{eq:SK}): 
\begin{equation}\label{eq:mono}
\frac{\int_{\lambda_{i}}^{\lambda_{f}}f_\lambda\,\lambda\,T_\zeta\,\rm{d}\lambda}
{\int_{\lambda_{i}}^{\lambda_{f}}\lambda\,T_\zeta\,\rm{d}\lambda} = 
10^{-0.4\,(m_{\zeta,{\tt SM}}-\epsilon_\zeta)}\,c\,f_\nu^{0}\,G(\lambda)
\end{equation}
where 
\begin{equation}
G(\lambda) = \frac{\int_{\lambda_{i}}^{\lambda_{f}}\frac{T_\zeta}{\lambda}\,\rm{d}\lambda}{\int_{\lambda_{i}}^{\lambda_{f}}\lambda\,T_\zeta\,\rm{d}\lambda}.
\end{equation}
Similarly, the in-band flux is: 
\begin{equation}\label{eq:band}
\int_{\lambda_{i}}^{\lambda_{f}} f_\lambda\,\lambda\,T_\zeta\,{\rm d}\lambda = 
10^{-0.4\,(m_{\zeta,{\tt SM}}-\epsilon_\zeta)}\,c\,f_\nu^{0}\,H(\lambda)
\end{equation}
where 
\begin{equation}
H(\lambda) = Bw(\lambda)\,G(\lambda) = \int_{\lambda_{i}}^{\lambda_{f}}\frac{T_\zeta}{\lambda}\,\rm{d}\lambda,
\end{equation}
and $Bw(\lambda)$ is the bandwidth. Monochromatic fluxes are associated with
isophotal wavelengths, whose calculation is non-trivial because of
discontinuities in stellar spectra \citep[e.g.,][]{tv05,c06,rieke08}. The
effective wavelength $\lambda_{\rm{eff}}$ is thus a useful approximation
\citep[e.g.,][]{golay74}:
\begin{equation}
\lambda_{\rm{eff}} = \frac{\int_{\lambda_i}^{\lambda_f} \lambda^2 f_{\lambda} T_\zeta \rm{d}\lambda}{\int_{\lambda_i}^{\lambda_f} \lambda f_{\lambda} T_\zeta \rm{d}\lambda}
\end{equation}
Values for $\epsilon_\zeta$, $G(\lambda)$, $H(\lambda)$, $Bw(\lambda)$ and
$\lambda_{\rm{eff}}$ are listed in Table \ref{tab:zp}, where the CALSPEC
spectrum of Vega has been adopted to compute the effective wavelength. For
example, an object with $m_{g,{\tt SM}}=g=15$ will have a
monochromatic flux of  $4.261 \times 10^{-15}\,\rm{erg}\,\rm{s}^{-1}\,\rm{cm}^{-2} \rm{\AA}^{-1}$
and an in-band flux of $6.182 \times 10^{-12}\,\rm{erg}\,\rm{s}^{-1}\,\rm{cm}^{-2}$.

\bibliographystyle{aj}
\bibliography{refs}

\begin{thebibliography}{}

\bibitem[\protect\citeauthoryear{{Abolfathi} et~al.}{{Abolfathi}
  et~al.}{2018}]{dr14}
{Abolfathi}, B., et~al. 2018, \apjs, 235, 42

\bibitem[\protect\citeauthoryear{{Alonso}, {Arribas}, \&
  {Martinez-Roger}}{{Alonso} et~al.}{1996}]{alonso96:irfm}
{Alonso}, A., {Arribas}, S.,  \& {Martinez-Roger}, C. 1996, \aaps, 117, 227

\bibitem[\protect\citeauthoryear{{Arenou} et~al.}{{Arenou}
  et~al.}{2018}]{arenou18}
{Arenou}, F., et~al. 2018, ArXiv e-prints

\bibitem[\protect\citeauthoryear{{{\'A}rnad{\'o}ttir}, {Feltzing}, \&
  {Lundstr{\"o}m}}{{{\'A}rnad{\'o}ttir} et~al.}{2010}]{arnadottir10}
{{\'A}rnad{\'o}ttir}, A.~S., {Feltzing}, S.,  \& {Lundstr{\"o}m}, I. 2010,
  \aap, 521, A40

\bibitem[\protect\citeauthoryear{{Bessell} et~al.}{{Bessell}
  et~al.}{2011}]{bbs11}
{Bessell}, M., {Bloxham}, G., {Schmidt}, B., {Keller}, S., {Tisserand}, P.,  \&
  {Francis}, P. 2011, \pasp, 123, 789

\bibitem[\protect\citeauthoryear{{Bessell} \& {Murphy}}{{Bessell} \&
  {Murphy}}{2012}]{bm12}
{Bessell}, M.,  \& {Murphy}, S. 2012, \pasp, 124, 140

\bibitem[\protect\citeauthoryear{{Bessell}}{{Bessell}}{2005}]{bessell05}
{Bessell}, M.~S. 2005, \araa, 43, 293

\bibitem[\protect\citeauthoryear{{Blackwell}, {Lynas-Gray}, \&
  {Petford}}{{Blackwell} et~al.}{1991}]{blp91}
{Blackwell}, D.~E., {Lynas-Gray}, A.~E.,  \& {Petford}, A.~D. 1991, \aap, 245,
  567

\bibitem[\protect\citeauthoryear{{Blackwell} et~al.}{{Blackwell}
  et~al.}{1990}]{bpa90}
{Blackwell}, D.~E., {Petford}, A.~D., {Arribas}, S., {Haddock}, D.~J.,  \&
  {Selby}, M.~J. 1990, \aap, 232, 396

\bibitem[\protect\citeauthoryear{{Blackwell}, {Petford}, \&
  {Shallis}}{{Blackwell} et~al.}{1980}]{blackwell80}
{Blackwell}, D.~E., {Petford}, A.~D.,  \& {Shallis}, M.~J. 1980, \aap, 82, 249

\bibitem[\protect\citeauthoryear{{Blackwell}, {Shallis}, \&
  {Selby}}{{Blackwell} et~al.}{1979}]{blackwell79}
{Blackwell}, D.~E., {Shallis}, M.~J.,  \& {Selby}, M.~J. 1979, \mnras, 188, 847

\bibitem[\protect\citeauthoryear{{Boeche} et~al.}{{Boeche}
  et~al.}{2014}]{boeche14}
{Boeche}, C., et~al. 2014, \aap, 568, A71

\bibitem[\protect\citeauthoryear{{Bohlin}}{{Bohlin}}{2007}]{bo07}
{Bohlin}, R.~C. 2007, in Astronomical Society of the Pacific Conference Series,
  Vol. 364, The Future of Photometric, Spectrophotometric and Polarimetric
  Standardization, ed. C.~{Sterken}, 315

\bibitem[\protect\citeauthoryear{{Bohlin}}{{Bohlin}}{2014}]{bo14}
{Bohlin}, R.~C. 2014, \aj, 147, 127

\bibitem[\protect\citeauthoryear{{Bohlin}, {Dickinson}, \& {Calzetti}}{{Bohlin}
  et~al.}{2001}]{bdc01}
{Bohlin}, R.~C., {Dickinson}, M.~E.,  \& {Calzetti}, D. 2001, \aj, 122, 2118

\bibitem[\protect\citeauthoryear{{Buder} et~al.}{{Buder}
  et~al.}{2018}]{buder18}
{Buder}, S., et~al. 2018, \mnras, 478, 4513

\bibitem[\protect\citeauthoryear{{Cardelli}, {Clayton}, \& {Mathis}}{{Cardelli}
  et~al.}{1989}]{ccm89}
{Cardelli}, J.~A., {Clayton}, G.~C.,  \& {Mathis}, J.~S. 1989, \apj, 345, 245

\bibitem[\protect\citeauthoryear{{Casagrande}, {Portinari}, \&
  {Flynn}}{{Casagrande} et~al.}{2006}]{c06}
{Casagrande}, L., {Portinari}, L.,  \& {Flynn}, C. 2006, \mnras, 373, 13

\bibitem[\protect\citeauthoryear{{Casagrande} et~al.}{{Casagrande}
  et~al.}{2014a}]{c14b}
{Casagrande}, L., et~al. 2014a, \mnras, 439, 2060

\bibitem[\protect\citeauthoryear{{Casagrande} et~al.}{{Casagrande}
  et~al.}{2010}]{c10}
{Casagrande}, L., {Ram{\'{\i}}rez}, I., {Mel{\'e}ndez}, J., {Bessell}, M.,  \&
  {Asplund}, M. 2010, \aap, 512, A54

\bibitem[\protect\citeauthoryear{{Casagrande} et~al.}{{Casagrande}
  et~al.}{2011}]{c11}
{Casagrande}, L., {Sch{\"o}nrich}, R., {Asplund}, M., {Cassisi}, S.,
  {Ram{\'{\i}}rez}, I., {Mel{\'e}ndez}, J., {Bensby}, T.,  \& {Feltzing}, S.
  2011, \aap, 530, A138

\bibitem[\protect\citeauthoryear{{Casagrande} et~al.}{{Casagrande}
  et~al.}{2016}]{c16}
{Casagrande}, L., et~al. 2016, \mnras, 455, 987

\bibitem[\protect\citeauthoryear{{Casagrande} et~al.}{{Casagrande}
  et~al.}{2014b}]{c14a}
{Casagrande}, L., et~al. 2014b, \apj, 787, 110

\bibitem[\protect\citeauthoryear{{Casagrande} \& {VandenBerg}}{{Casagrande} \&
  {VandenBerg}}{2014}]{cv14}
{Casagrande}, L.,  \& {VandenBerg}, D.~A. 2014, \mnras, 444, 392

\bibitem[\protect\citeauthoryear{{Casagrande} \& {VandenBerg}}{{Casagrande} \&
  {VandenBerg}}{2018a}]{cv18b}
{Casagrande}, L.,  \& {VandenBerg}, D.~A. 2018a, \mnras, 479, L102

\bibitem[\protect\citeauthoryear{{Casagrande} \& {VandenBerg}}{{Casagrande} \&
  {VandenBerg}}{2018b}]{cv18}
{Casagrande}, L.,  \& {VandenBerg}, D.~A. 2018b, \mnras, 475, 5023

\bibitem[\protect\citeauthoryear{{Ciuc{\v a}} et~al.}{{Ciuc{\v a}}
  et~al.}{2018}]{ciuca18}
{Ciuc{\v a}}, I., {Kawata}, D., {Lin}, J., {Casagrande}, L., {Seabroke}, G.,
  \& {Cropper}, M. 2018, \mnras, 475, 1203

\bibitem[\protect\citeauthoryear{{Covey} et~al.}{{Covey}
  et~al.}{2007}]{covey07}
{Covey}, K.~R., et~al. 2007, \aj, 134, 2398

\bibitem[\protect\citeauthoryear{{Da Costa} et~al.}{{Da Costa}
  et~al.}{2018}]{dacosta}
{Da Costa}, G., et~al. 2018, in preparation

\bibitem[\protect\citeauthoryear{{De Silva} et~al.}{{De Silva}
  et~al.}{2015}]{galah}
{De Silva}, G.~M., et~al. 2015, \mnras, 449, 2604

\bibitem[\protect\citeauthoryear{{Doi} et~al.}{{Doi} et~al.}{2010}]{doi10}
{Doi}, M., et~al. 2010, \aj, 139, 1628

\bibitem[\protect\citeauthoryear{{Eisenstein} et~al.}{{Eisenstein}
  et~al.}{2006}]{eisen06}
{Eisenstein}, D.~J., et~al. 2006, \apjs, 167, 40

\bibitem[\protect\citeauthoryear{{Fitzpatrick}}{{Fitzpatrick}}{1999}]{fitz99}
{Fitzpatrick}, E.~L. 1999, \pasp, 111, 63

\bibitem[\protect\citeauthoryear{{Francis} \& {Wills}}{{Francis} \&
  {Wills}}{1999}]{fw99}
{Francis}, P.~J.,  \& {Wills}, B.~J. 1999, in Astronomical Society of the
  Pacific Conference Series, Vol. 162, Quasars and Cosmology, ed. G.~{Ferland}
  \& J.~{Baldwin}, 363

\bibitem[\protect\citeauthoryear{{Fukugita} et~al.}{{Fukugita}
  et~al.}{1996}]{fuku96}
{Fukugita}, M., {Ichikawa}, T., {Gunn}, J.~E., {Doi}, M., {Shimasaku}, K.,  \&
  {Schneider}, D.~P. 1996, \aj, 111, 1748

\bibitem[\protect\citeauthoryear{{Gaia Collaboration} et~al.}{{Gaia
  Collaboration} et~al.}{2018}]{Gaia18}
{Gaia Collaboration}, {Brown}, A.~G.~A., {Vallenari}, A., {Prusti}, T., {de
  Bruijne}, J.~H.~J., {Babusiaux}, C.,  \& {Bailer-Jones}, C.~A.~L. 2018, ArXiv
  e-prints

\bibitem[\protect\citeauthoryear{{Golay}}{{Golay}}{1974}]{golay74}
{Golay}, M., ed. 1974, Astrophysics and Space Science Library, Vol.~41,
  {Introduction to astronomical photometry}

\bibitem[\protect\citeauthoryear{{Henden} et~al.}{{Henden}
  et~al.}{2016}]{apass2016}
{Henden}, A.~A., {Templeton}, M., {Terrell}, D., {Smith}, T.~C., {Levine}, S.,
  \& {Welch}, D. 2016, VizieR Online Data Catalog, 2336

\bibitem[\protect\citeauthoryear{{High} et~al.}{{High} et~al.}{2009}]{high09}
{High}, F.~W., {Stubbs}, C.~W., {Rest}, A., {Stalder}, B.,  \& {Challis}, P.
  2009, \aj, 138, 110

\bibitem[\protect\citeauthoryear{{Holberg} \& {Bergeron}}{{Holberg} \&
  {Bergeron}}{2006}]{hb06}
{Holberg}, J.~B.,  \& {Bergeron}, P. 2006, \aj, 132, 1221

\bibitem[\protect\citeauthoryear{{Howes} et~al.}{{Howes}
  et~al.}{2016}]{howes16}
{Howes}, L.~M., et~al. 2016, \mnras, 460, 884

\bibitem[\protect\citeauthoryear{{Ivezi{\'c}}, {Beers}, \&
  {Juri{\'c}}}{{Ivezi{\'c}} et~al.}{2012}]{ibj}
{Ivezi{\'c}}, {\v Z}., {Beers}, T.~C.,  \& {Juri{\'c}}, M. 2012, \araa, 50, 251

\bibitem[\protect\citeauthoryear{{Ivezi{\'c}} et~al.}{{Ivezi{\'c}}
  et~al.}{2008}]{ivezic08}
{Ivezi{\'c}}, {\v Z}., et~al. 2008, \apj, 684, 287

\bibitem[\protect\citeauthoryear{{Ivezi{\'c}} et~al.}{{Ivezi{\'c}}
  et~al.}{2007}]{ivezic07}
{Ivezi{\'c}}, {\v Z}., et~al. 2007, \aj, 134, 973

\bibitem[\protect\citeauthoryear{{Karovicova} et~al.}{{Karovicova}
  et~al.}{2018}]{karo}
{Karovicova}, I., et~al. 2018, \mnras, 475, L81

\bibitem[\protect\citeauthoryear{{Keller} et~al.}{{Keller}
  et~al.}{2014}]{keller14}
{Keller}, S.~C., et~al. 2014, \nat, 506, 463

\bibitem[\protect\citeauthoryear{{Keller} et~al.}{{Keller}
  et~al.}{2007}]{keller07}
{Keller}, S.~C., et~al. 2007, \pasa, 24, 1

\bibitem[\protect\citeauthoryear{{Kunder} et~al.}{{Kunder}
  et~al.}{2017}]{kunder}
{Kunder}, A., et~al. 2017, \aj, 153, 75

\bibitem[\protect\citeauthoryear{{Lallement} et~al.}{{Lallement}
  et~al.}{2003}]{lallement03}
{Lallement}, R., {Welsh}, B.~Y., {Vergely}, J.~L., {Crifo}, F.,  \& {Sfeir}, D.
  2003, \aap, 411, 447

\bibitem[\protect\citeauthoryear{{MacDonald} et~al.}{{MacDonald}
  et~al.}{2004}]{macdonald04}
{MacDonald}, E.~C., et~al. 2004, \mnras, 352, 1255

\bibitem[\protect\citeauthoryear{{McClure}}{{McClure}}{1976}]{mcclure76}
{McClure}, R.~D. 1976, \aj, 81, 182

\bibitem[\protect\citeauthoryear{{Ness} et~al.}{{Ness} et~al.}{2015}]{cannon}
{Ness}, M., {Hogg}, D.~W., {Rix}, H.-W., {Ho}, A.~Y.~Q.,  \& {Zasowski}, G.
  2015, \apj, 808, 16

\bibitem[\protect\citeauthoryear{{Nordstr{\"o}m} et~al.}{{Nordstr{\"o}m}
  et~al.}{2004}]{nordstrom04}
{Nordstr{\"o}m}, B., et~al. 2004, \aap, 418, 989

\bibitem[\protect\citeauthoryear{{O'Donnell}}{{O'Donnell}}{1994}]{od94}
{O'Donnell}, J.~E. 1994, \apj, 422, 158

\bibitem[\protect\citeauthoryear{{Rieke} et~al.}{{Rieke}
  et~al.}{2008}]{rieke08}
{Rieke}, G.~H., et~al. 2008, \aj, 135, 2245

\bibitem[\protect\citeauthoryear{{Schlafly} et~al.}{{Schlafly}
  et~al.}{2016}]{sms16}
{Schlafly}, E.~F., et~al. 2016, \apj, 821, 78

\bibitem[\protect\citeauthoryear{{Schlegel}, {Finkbeiner}, \&
  {Davis}}{{Schlegel} et~al.}{1998}]{sfd98}
{Schlegel}, D.~J., {Finkbeiner}, D.~P.,  \& {Davis}, M. 1998, \apj, 500, 525

\bibitem[\protect\citeauthoryear{{Skrutskie} et~al.}{{Skrutskie}
  et~al.}{2006}]{skr06}
{Skrutskie}, M.~F., et~al. 2006, \aj, 131, 1163

\bibitem[\protect\citeauthoryear{{Stetson}}{{Stetson}}{2005}]{Stetson05}
{Stetson}, P.~B. 2005, \pasp, 117, 563

\bibitem[\protect\citeauthoryear{{Str{\"o}mgren}}{{Str{\"o}mgren}}{1951}]{stro51}
{Str{\"o}mgren}, B. 1951, \aj, 56, 142

\bibitem[\protect\citeauthoryear{{Taylor}}{{Taylor}}{2005}]{topcat}
{Taylor}, M.~B. 2005, in Astronomical Society of the Pacific Conference Series,
  Vol. 347, Astronomical Data Analysis Software and Systems XIV, ed.
  P.~{Shopbell}, M.~{Britton}, \& R.~{Ebert}, 29

\bibitem[\protect\citeauthoryear{{Tokunaga} \& {Vacca}}{{Tokunaga} \&
  {Vacca}}{2005}]{tv05}
{Tokunaga}, A.~T.,  \& {Vacca}, W.~D. 2005, \pasp, 117, 421

\bibitem[\protect\citeauthoryear{{White} et~al.}{{White} et~al.}{2018}]{twhite}
{White}, T.~R., et~al. 2018, \mnras, 477, 4403

\bibitem[\protect\citeauthoryear{{Wolf} et~al.}{{Wolf} et~al.}{2018}]{wolf18}
{Wolf}, C., et~al. 2018, \pasa, 35, e010

\bibitem[\protect\citeauthoryear{{Yong} et~al.}{{Yong} et~al.}{2013}]{yong13}
{Yong}, D., et~al. 2013, \apj, 762, 27

\bibitem[\protect\citeauthoryear{{Yuan} et~al.}{{Yuan} et~al.}{2015}]{yuan15}
{Yuan}, H., {Liu}, X., {Xiang}, M., {Huang}, Y., {Zhang}, H.,  \& {Chen}, B.
  2015, \apj, 799, 133

\end{thebibliography}

\end{document}